\documentclass[12pt]{article}
\usepackage[utf8]{inputenc}

\usepackage{fullpage}

\usepackage{amsfonts,amsmath,amssymb,amsthm,bbm}
\usepackage{float,mathtools}
\usepackage[shortlabels]{enumitem}
\usepackage{varwidth}

\usepackage{graphicx}
\usepackage[ruled]{algorithm}
\usepackage{algpseudocode}
\usepackage[numbers]{natbib}
\usepackage[colorlinks]{hyperref}
\usepackage{subfig}

\newenvironment{proof*}{\paragraph{Proof:}}{\hfill$\blacksquare$}

\newtheorem{theorem}{Theorem}

\newtheorem*{theorem*}{Theorem}

\newtheorem{proposition}{Proposition}

\theoremstyle{definition}

\newtheorem{setting}{Setting}
\newtheorem{remark}{Remark}

\newtheorem{assumption}{Assumption}

\title{Conformal Prediction for Network-Assisted Regression}
\author{Robert Lunde\thanks{Department of Mathematics and Statistics, Washington University in St. Louis} \and  Elizaveta Levina \thanks{Department of Statistics, University of Michigan} \and  Ji Zhu \footnotemark[2]}

\begin{document}
\maketitle
\abstract{An important problem in  network analysis is predicting a node attribute using both network covariates, such as graph embedding coordinates or local subgraph counts, and conventional node covariates, such as demographic characteristics. While standard regression methods that make use of both types of covariates may be used for prediction, statistical inference is complicated by the fact that the nodal summary statistics are often dependent in complex ways.   We show that under a mild joint exchangeability assumption, a network analog of conformal prediction achieves finite sample validity for a wide range of network covariates.  We also show that a form of  asymptotic conditional validity is achievable.  The methods are illustrated on both simulated networks and a citation network dataset.}  

\section{Introduction}    
Network analysis has emerged as a key framework for studying behavior in many fields, including sociology \citep{doi:10.1177/0038038588022001007}, biology \citep{WEY2008333}, economics \citep{JACKSON2011511}, and public health \citep{doi:10.1146/annurev-publhealth-031816-044528}.  A common task in these disciplines is predicting some attribute $Y$ of an individual in the network using both conventional covariates $X$, such as sex or income, along with network covariates, such as degrees or the individual's ``position" in the network. For example, education researchers may be interested in predicting academic achievement, which is known to depend in part on the individual's social network (e.g. \citet{Stadtfeld792}).  As another example, tech companies may be interested in predicting user engagement based on the behavior of the user's social circle. 
 
These questions can be formulated as prediction problems on data connected by a network; a plethora of tools, from linear regression to deep neural networks, may be used to generate predictions.  However, statistical \textit{inference}, that is, attaching a measure of uncertainty to the predictions or other estimated quantities, is complicated by the fact that networks typically induce non-standard dependence structure among observations.  Consider for example node degrees, which are one of the simplest and most widely used network statistics, and may be a useful covariate for making some prediction about the node.  The degrees of two nodes in the network are generally dependent since they share one possible edge, and can more generally jointly depend on some underlying propensities to form edges.
 
 In other settings, conformal prediction, pioneered by Vovk and colleagues in the 1990s (see literature review in Section \ref{sec-related-work}), has emerged as a powerful and flexible tool for quantifying uncertainty associated with modern regression methods.  When the training pairs $(Y_1, X_1), \ldots, (Y_{n}, X_{n})$ are exchangeable, conformal prediction offers confidence sets  $\hat{C}_n$  such that for a new data point $(Y_{n+1},X_{n+1})$ and pre-specified level $\alpha$, we have the finite-sample guarantee, 
 \begin{align}
 \label{eq-conformal-prediction-guarantee}
P(Y_{n+1} \in \hat{C}_n(X_{n+1})) \geq 1-\alpha.
 \end{align} 
 
Most applications of conformal prediction deal with the case where the pairs are i.i.d.\ and do not take full advantage of the generality of the theory.  In the network setting, however,   leveraging exchangeability will be crucial for valid conformal prediction.

The idea of exchangeability has played a pivotal role in the development of statistical models for network data.  In many situations, it is natural to posit that the probability distribution associated with the network is invariant to relabeling (or permuting) the nodes.  Many commonly used network models satisfy this assumption, including the stochastic block model (SBM) and its extensions \citep{holland-sbm,airoldi-mmsb,karrer-newman-dcsbm}, random dot product graphs (RDPG) \citep{young-schneiderman-rdpg,rubindelanchy2020statistical}, latent space models \citep{hoff-raftery-handcock-latent-space-model}, and graphons \citep{LOVASZ2006933,Bickel-Chen-on-modularity,borgs-lp-part-one}. 

While vertex (joint) exchangeability is a natural notion for networks, it does not immediately imply the validity of conformal prediction methods for such data.  A key question is: under what conditions are covariates computed from such a network exchangeable?  Surprisingly, it turns out that network covariates are \textit{always exchangeable} provided they satisfy a mild symmetry condition that is closely related to vertex exchangeability.    One of our main contributions is establishing the validity of conformal prediction for network-assisted prediction; we are not aware of any other methods available for statistical inference for regression models incorporating information from any vertex exchangeable graph.   

Our second major contribution is establishing a form of asymptotic conditional validity for network regression.  While the property \eqref{eq-conformal-prediction-guarantee} serves as the primary theoretical justification for conformal prediction, for certain applications one may want a prediction interval with a stronger conditional guarantee:
\begin{align}
 \label{eq-conformal-prediction-conditional-guarantee}
P(Y_{n+1} \in \hat{C}_n(X_{n+1}) \ | \ X_{n+1} = x) \geq 1-\alpha \ \ a.s.
 \end{align} 
Such a guarantee would ensure that the conformal prediction sets have the desired coverage level conditionally even for ``difficult" cases, which may be of the most practical interest to begin with.  Unfortunately, \eqref{eq-conformal-prediction-conditional-guarantee} is unachievable in finite samples over general classes of probability distributions \citep{pmlr-v25-vovk12,https://doi.org/10.1111/rssb.12021,10.1093/imaiai/iaaa017}.  Recently, \citet{Chernozhukove2107794118} showed that a form of asymptotic conditional validity is achievable asymptotically so long as the conditional distribution function $F_{Y|X}$ can be consistently estimated. We will show that in the network setting consistent estimation is possible and therefore a form of conditional validity is attainable.

The rest of the paper is organized as follows.  In Section \ref{sec-related-work}, we provide a literature review on conformal prediction and inference for network regression.  In Section \ref{sec:problem-setup-notation}, we introduce relevant notation and background, and describe the conformal prediction algorithm in the network prediction setting.   We state our main results in Section \ref{sec:main-results}.  Empirical results on both  simulated and real data are presented in Section \ref{sec:experiments}.   Section \ref{sec:discussion} concludes with discussion.  

\section{Related Work}
\label{sec-related-work}

\subsection{Conformal Prediction}


Distribution-free uncertainty quantification has received substantial attention in recent years and the conformal prediction literature is growing rapidly. The core theory for conformal prediction is laid out in  \citet{vovk-algorithmic-world}; see \citet{angelopoulos2021gentle} for a historical account.    The procedure was popularized in statistics by  \citet{doi:10.1080/01621459.2012.751873} and \citet{JMLR:v9:shafer08a} and further studied by \citet{pmlr-v25-vovk12}, \citet{https://doi.org/10.1111/rssb.12021}, \citet{doi:10.1080/01621459.2017.1307116}, and others.  

Two important recent directions have been the study of conditional coverage and establishing validity of conformal prediction in novel settings. For the former, \citet{https://doi.org/10.1111/rssb.12021, 10.1093/imaiai/iaaa017} consider enlargements of the conditioning set $\{X= x\}$ to $\{X \in A_k \}$ for a finite partition $\bigcup_{k=1}^K A_k$ of the ambient space $\mathcal{X}$.  \citet{https://doi.org/10.1111/rssb.12021} further show that asymptotic conditional validity is achievable using nonparametric density estimation.  In Section \ref{sec:conditional-validity}, we consider asymptotic conditional validity for networks when distribution regression is used; analogous procedures are studied in other settings by \citet{pmlr-v108-izbicki20a} and \citet{Chernozhukove2107794118}.   

Establishing the validity of conformal prediction in new settings often involves showing that \eqref{eq-conformal-prediction-guarantee} holds asymptotically even when exchangeability is violated. Along these lines, \citet{DBLP:conf/colt/ChernozhukovWZ18} and \citet{Chernozhukove2107794118} consider time series with weak dependence,  \citet{https://doi.org/10.1111/rssb.12445} consider survival analysis and \citet{candes2021conformalized} consider individual treatment effect estimation in causal inference.  
The validity of a weighted conformal prediction procedure for covariate shift problems was established in \citet{NEURIPS2019_8fb21ee7}. Other settings in which conformal prediction has been studied include functional data \citep{lei-functional-data}, random effects models \citep{dunn2020distribution}, and ranking \citep{cauchois2022predictive}.

In the network setting, \citet{https://doi.org/10.48550/arxiv.2109.1272} consider conformal prediction for detection of anomalous edges for edge-exchangeable random graphs.  Edge exchangeability can be a useful framework for modeling randomly sampled interactions \citep{doi:10.1080/01621459.2017.1341413}, while vertex exchangeability remains the most natural assumption for many network settings. The main criticism of the vertex exchangeability assumption is that it does not allow for features like sparsity in the limit;  however,  we only require a form of finite exchangeability and thus do not directly deal with embedding our problem in an asymptotic representation.  Moreover, our work allows general classes of network covariates, which are not considered in \citet{https://doi.org/10.48550/arxiv.2109.1272}.           

\subsection{Network-Assisted Prediction}
Despite its practical importance, prediction on network-linked data has only recently started to receive attention in statistics:  for example, \citet{https://doi.org/10.1002/jae.2426},   \citet{10.1214/16-AOS1476} and \citet{doi:10.1080/01621459.2019.1617153} all consider models in which the covariates include neighborhood effects.  \citet{10.1214/18-AOAS1205} develop prediction models that are fit using a cohesion penalty, which enforces smoothly varying predictions over network neighborhoods.   \citet{le2021linear}  generalize the cohesion penalty approach to include the case when the intercept vector lies in the span of a low rank matrix.  \citet{pmlr-v139-mao21a} consider nonparametric regression using network covariates and establish consistency of kernel regression under a smooth graphon assumption when certain network covariates are used.  In economics, some approaches based on using neighborhood statistics as covariates  have been considered by \citet{10.2307/2298123, LEE2007333, BRAMOULLE200941}.      

In machine learning, deep neural networks have been used for graph-structured data.   Early work includes \citet{1555942} and \citet{4700287}, and a comprehensive review can be found in  \citet{GOYAL201878}.  A common approach to deep learning with networks involves computing low-dimensional embeddings for nodes, or in other words fitting a latent variable model, and then using the latent variables as features.    Some popular embedding approaches sample neighborhoods of nodes using random walks, such as node2vec \citep{10.1145/2939672.2939754} and DeepWalk \citep{10.1145/2623330.2623732}. Approaches based on deep autoencoders produce embeddings in which ``similar" nodes are close in the embedding space, such as \citep{10.1145/2939672.2939753,10.5555/3015812.3015982}.   We will focus on embeddings motivated by statistical models, but many other embeddings used in the deep learning literature satisfy the mild regularity conditions needed for unconditional validity; see Section \ref{sec:unconditional-validity} for details. 
  
\section{Problem Setup and a Conformal Prediction Algorithm}
\label{sec:problem-setup-notation}
\subsection{Jointly Exchangeable Models}
We start from definitions and notation.   Let $Y_1, \ldots, Y_{2n+1} \in \mathbb{R}$ denote the response variables, $X_1, \ldots, X_{2n+1} \in \mathbb{R}^d$ denote covariates, and let $A$ denote a corresponding $(2n+1) \times (2n+1)$ connection matrix, where $A_{ij}$ provides information about the relationship between nodes $i$ and $j$.  
Typically $A$ will be a binary symmetric adjacency matrix, but weighted and directed graphs are permitted.  One may also consider a collection of connection matrices, but we stick to one such matrix for simplicity. Since our main focus will be on split conformal inference, we use a sample size of $2n+1$ rather than $n+1$ for notational convenience.  

Now for $1\leq i, j \leq 2n+1$, let $V_{ij} = (Y_i,Y_j, X_i, X_j, A_{ij})$.   Let $\sigma:[2n+1] \mapsto [2n+1]$ denote a permutation function and $\stackrel{d}{=}$ denote equality in distribution.  We make the following assumption:
\begin{assumption}
\label{assumption-exchangeability}
The array $(V_{ij})_{1 \leq i, j \leq 2n+1}$ is jointly exchangeable; that is, for any permutation function $\sigma$,
\begin{align}
(V_{\sigma(i) \sigma(j)})_{1 \leq i,j \leq 2n+1} \stackrel{d}{=} (V_{ij})_{1 \leq i, j \leq 2n+1},
\end{align}
\end{assumption}

Assumption \ref{assumption-exchangeability} is very general and includes multiple commonly used and natural network models.   We describe two important settings that satisfy this assumption next.  

\begin{setting}[Independent Triplets and a Graphon Model]
\label{setting-independent-triples}
 Suppose that $(Y_1, X_1,\xi_1),$ $\ldots, (Y_{2n+1}, X_{2n+1}, \xi_{2n+1})$ are i.i.d.\ triplets, where $\xi_1,\ldots, \xi_{2n+1}$ are latent positions marginally uniformly distributed on $[0,1]$, and the adjacency matrix is generated as 
  \begin{align}
 \label{eq:sparse-graphon-model}
 A_{ij} = A_{ji} =  \mathbbm{1}(\eta_{ij} \leq \rho_n w(\xi_i, \xi_j) \wedge 1 ) .  
 \end{align}
Here $\{\eta_{ij} \}_{1\leq i<j \leq 2n+1}$ is another set of i.i.d.\ $\mathrm{Uniform}[0,1]$ variables independent from all other random variables, $\rho_n$ controls sparsity of the network, and $w$ is a non-negative function symmetric in its arguments which satisfies $\int_0^1 \int_0^1 w(u,v) \ du \ dv =1$.  While it is immaterial to the present work, we assume no self-loops, i.e.,  $A_{ii}= 0$ for all $i$.  This model, originally due to \citet{aldous-representation-array} and \citet{hoover-exchangeability} and now known as the sparse graphon model \citep{klopp-oracle-inequalities,10.1214/19-STS736}, was first considered in statistics by \citet{Bickel-Chen-on-modularity}. 
As $n$ grows, it is natural to focus on the case $\rho_n \rightarrow 0$, since  most real world graphs are sparse, in the sense that they have $o(n^2)$ edges. 

Note that since the triplets are independent, dependence arising from linked nodes is not directly modeled in this setup.  As discussed in Example \ref{example-neighbor-average} below, statistics such as neighborhood averages may be included as a covariate in the fitted model, but are viewed as estimates of node-level quantities in this setup.  We believe that this viewpoint is appropriate for snapshots from a large network, where statistics such as edge-weighted covariates reflect information about the individual in question in terms of the types of people they associate with or their shared preferences and are robust to the actions or beliefs of a random acquaintance.
\end{setting}
Alternatively, one could consider the following setting where the response depends explicitly on neighborhood averages.  

\begin{setting}[Regression with Neighborhood and Node Effects]
\label{setting-neighborandnode}
Suppose that for some $N \geq 2n+1$, which represents the number of nodes in the population, $(X_1, \xi_1,\epsilon_1) ,\ldots,$ $(X_{N}, \xi_{N},\epsilon_{N})$ are exchangeable and that $A$ is generated by the sparse graphon model  \eqref{eq:sparse-graphon-model}.   Let $\alpha_{ij}^{(k)}$ be a binary random variable equal to 1 if the shortest path from node $i$ to $j$ is of length $k$, and 0 otherwise.  Let 
\begin{align*}
\widetilde{D}_i^{(k)} = \sum_{j \neq i} \alpha_{ij}^{(k)}, \ \ 
\widetilde{X}_i^{(k)} = \frac{1}{  \widetilde{D}_i^{(k)}} \sum_{j \neq i} \alpha_{ij}^{(k)} X_j.
\end{align*}
Furthermore, let $\beta_{ij}$ is a weight function depending only on the length of the shortest path between nodes $i$ and $j$, and define the neighbor-weighted response:
\begin{align*}
\widetilde{Y}_i = \frac{1}{\sum_{j \neq i} \beta_{ij}} \sum_{j \neq i} \beta_{ij} Y_j
\end{align*}
Now, suppose that $Y_i$ may be represented as: 
\begin{align*}
Y_i = f(X_i, \xi_i, \widetilde{Y}_i, \widetilde{D}_i^{(1)}, \ldots, \widetilde{D}_i^{(2n)},  \widetilde{X}_i^{(1)}, \ldots, \widetilde{X}_i^{(2n)},\epsilon_i), 
\end{align*}
where $f$ is measurable function such that a unique solution exists almost surely.
This model is a nonparametric generalization of spatial autoregressive models for networks studied in, for example, \cite{10.2307/2298123}.  The model allows $Y_i$ to depend on averages of covariates over $k$-neighborhoods, with the natural assumption being that as $k$ increases, the influence of these nodes on $Y_i$ diminishes.  When $\widetilde{Y}_i$ is included in the model, the value of $Y_i$ is determined endogenously.  Moreover, when the model is linear in $\widetilde{Y}_i$, conditions for existence and uniqueness of a solution can be stated in terms of invertibility of an appropriate matrix. In either case, it may not be immediately obvious that this data generating process satisfies Assumption \ref{assumption-exchangeability}; the following proposition formally states that it does.  The proof can be found in the Appendix:
\end{setting}
\begin{proposition}
\label{prop-setting-exchangeable}
The data generating processes defined in Settings~1~and~2 satisfy Assumption \ref{assumption-exchangeability}.
\end{proposition}
    
\subsection{Network Covariates}
In both Settings \ref{setting-independent-triples} and \ref{setting-neighborandnode}, the latent positions $\xi_1, \ldots, \xi_{2n+1}$ are unobservable.  Instead, we fit a regression model to triplets of the form $(Y_i, X_i, \hat{Z}_i)$, 
where $\hat{Z}_i$ are local network statistics corresponding to node $i$; for concreteness, let $\hat{Z}_i \in \mathbb{R}^p$.  We allow $\hat{Z}_i$ to depend on both $(X_1, ,\ldots, X_{2n+1})$ and $A$.   Typically, the statistic $\hat{Z}_i$ may be viewed as an  estimate of a population quantity $Z_i = g(\xi_i)$ for some measurable $g$. 
We provide examples of such statistics below, which are by no means exhaustive.
\paragraph{Example 1:  Degrees. }
As discussed previously, a node's degree $D_i = \sum_{j \neq i} A_{ij}$ is a widely used and informative statistic.  In this case, a corresponding population-level quantity estimated by $D_i/(n-1)$ is $\mathbb{E}[D_i/(n-1) \ | \ \xi_i] = \int_0^1 \rho_n w(\xi_i,v) \ dv$.  Other local count statistics such as rooted stars and triangles involving a given node also fall into this framework.

\paragraph{Example 2: Generalized RDPG Embedding Coordinates. }
\label{example-ase}

Suppose that the graphon admits the spectral decomposition
\begin{align*}
w(\xi_i,\xi_j) = \sum_{r=1}^p \lambda_r \phi_r(\xi_i) \phi_r(\xi_j) - \sum_{s=1}^q \gamma_s \psi_s(\xi_i) \psi_s(\xi_j),    
\end{align*}
where $p$ and $q$ are integers, $\lambda_1 \ge \cdots \ge \lambda_p > 0 $ are positive eigenvalues, $\gamma_1 \ge \cdots \ge \gamma_q > 0$ are the magnitudes of the negative eigenvalues, and $\{\phi_r\}_{1 \leq r \leq p}$, $\{\psi_s\}_{1 \leq s \leq q }$ are the corresponding eigenfunctions of the operator $Tf = \int_0^1 w(u,v) f(v) \ dv$. 
Then the graphon can be represented as a difference of inner products \citep{10.1214/20-AOS1976},\
\begin{align*}
P(A_{ij} = 1  \ | \  U_i,V_i,U_j,V_j) = \langle U_i,U_j \rangle - \langle V_i, V_j \rangle ,   
\end{align*}
where 
\begin{align*}
U_{ir} = \lambda_r^{1/2} \phi_r(\xi_i), \ \ V_{is} = \gamma_s^{1/2} \psi_s(\xi_i) . 
\end{align*}

This inner product model is known as a generalized random dot product graph \citep{rubindelanchy2020statistical}. The most natural estimate of the latent positions  $\{(U_i$, $V_i)\}_{1 \leq i \leq 2n+1}$ is given by the adjacency spectral embedding \citep{doi:10.1080/01621459.2012.699795}, which estimates eigenvalues and eigenfunctions from the singular value decomposition of the adjacency matrix.  That is, let $\hat{\lambda}_1, \ldots, \hat{\lambda}_p$ and $\hat{\gamma}_1, \ldots, \hat{\gamma}_q$ denote the magnitudes of the $p$ largest positive eigenvalues and $q$ smallest negative eigenvalues of $A$ respectively, (according to above, $\hat{\gamma}$s are supposed to be positive) and suppose that the corresponding eigenvectors are given by $\{u_r\}_{1 \leq r \leq p}$ and $\{v_s\}_{1 \leq s \leq p}$, respectively.  

Then the latent positions are estimated by
\begin{align*}
\hat{U}_{ir} = \hat{\lambda}_r^{1/2} u_{ri}, \ \ \hat{V}_{is} = \hat{\gamma}_s^{1/2} v_{si}.   
\end{align*}
While the underlying latent positions $\{(U_i, V_i)\}_{1 \leq i \leq 2n+1}$ can only be inferred up to an unknown indefinite orthogonal rotation, rates of convergence in the $2 \rightarrow \infty$ norm after such a rotation have been established in \citet{rubindelanchy2020statistical}.  It should also be noted that even when the rank is infinite, a truncated version of such an embedding may be considered following \citet{10.1214/20-AOS1976}.

\paragraph{Example 3: Neighborhood Averages.}
\label{example-neighbor-average}
Let:
\begin{align*}
\hat{Z}_i = \frac{1}{D_i} \sum_{j\neq i} A_{ij} X_j   
\end{align*}
denote the average value of a covariate among node $i$'s neighbors.  This average can be viewed as an estimate of the population quantity 
\begin{align*}
Z_i = \frac{\mathbb{E}[ w(\xi_i, \xi_j) X_j \ | \ \xi_i ]}{\mathbb{E}[w(\xi_i, \xi_j) \ | \ \xi_i]}.
\end{align*}
One may also consider averages within a $k$-hop neighborhood for $k>1$.  
For the moment, we will not place any formal assumptions on the nature of the summary statistic $\hat{Z}_i$.  In Section \ref{sec:unconditional-validity}, we will see that the network statistics must treat the nodes symmetrically for conformal prediction to generalize to network data. 

\subsection{Split Conformal Prediction}

We start from some additional notation.  Let $\mathcal{D}_1 = \{(Y_1,X_1,\hat{Z}_1), \ldots, (Y_n,X_n,\hat{Z}_n)\}$ denote the training set and $\mathcal{D}_2 = \{(Y_{n+1},X_{n+1},\hat{Z}_{n+1}), \ldots, (Y_{2n},X_{2n},\hat{Z}_{2n})\}$ denote the validation set. While other choices are possible for the relative sizes of $\mathcal{D}_1$ and $\mathcal{D}_2$, we only consider this split for concreteness.  

Let $s(y,x,z \ ; \mathcal{D}_1): \mathbb{R} \times \mathbb{R}^d \times \mathbb{R}^p \mapsto \mathbb{R}^+$ denote a nonconformity score function, which measures how unusual a given triplet is.  With split conformal prediction, the nonconformity score is allowed to depend on, for example, a model trained on $\mathcal{D}_1$.  A common nonconformity score for regression problems is the  absolute residual $|y - \hat{\mu}_n(x,z)|$. While many other choices are possible and appropriate for certain applications, we will also consider a more generic score function $|1/2 - \hat{F}_{Y|X,Z}(y \ | \ x,z)|$, which measures how unusual that the value $y$ is relative to the empirical estimate of the distribution of $Y$ conditional on $X =x, Z = z$, and under some conditions yields asymptotic conditional validity.      

The split conformal procedure is a variant of conformal prediction that uses sample splitting.  The main advantage of the split conformal method is reduced computation.  The original conformal prediction procedure typically requires refitting the model for each grid point to construct a confidence region.  In contrast, with split conformal prediction, one fits the model only once.  The general split conformal procedure we propose for prediction problems on network-linked data is presented in Algorithm 1.   
\begin{algorithm}
\caption{Split Conformal Prediction}
\hspace*{\algorithmicindent} \textbf{Input:} Data $(Y_1,X_1), \ldots, (Y_{2n},X_{2n})$, $(2n+1)\times(2n+1)$ adjacency matrix $A$, level $\alpha$, new point $X_{2n+1}$. \\ 
\hspace*{\algorithmicindent}
\textbf{Output:} Confidence Set $\widehat{C}_n$
\begin{algorithmic}[1]
\State Construct summary statistics $\hat{Z}_1, \ldots, \hat{Z}_{2n+1}$ from $(A,X)$.
\State Split into two folds $\mathcal{D}_1 = (Y_i,X_i,\hat{Z}_i )_{1\leq i \leq n}$, $\mathcal{D}_2 = (Y_i,X_i,\hat{Z}_i)_{n+1 \leq i \leq 2n}$.
\For{ $i \in \{n+1, \ldots, 2n\}$ }
\State  $S_{i} \leftarrow s(Y_i,X_i, \hat{Z}_i \ ; \ \mathcal{D}_1)$
\EndFor
\State $d \leftarrow  (1-\alpha)(1+\frac{1}{n}) $ empirical quantile of $\{S_i\}_{n+1 \leq i \leq 2n}$
\State $\widehat{C}_n \leftarrow \{y: s(y,X_{2n+1}, \widehat{Z}_{2n+1}) \leq d \}$
\end{algorithmic}
\end{algorithm}

\section{Main Results: Validity of Conformal Prediction with Network Data}
\label{sec:main-results}

\subsection{Finite-Sample Unconditional Validity}
\label{sec:unconditional-validity}

Our first main result, the validity of conformal prediction for network regression, is based on the intuition that natural network summary statistics exhibit certain symmetry properties and are thus exchangeable.  
To rigorously flesh out this idea, we first establish a modest generalization
of a theorem of \citet{dean-verducci} and \citet{commenges-transformations} stated in the recent review article of \citet{kuchibhotla2021exchangeability}.  
\begin{proposition}
\label{theorem-exchangeability-transformations}
 Let $X$ be a random variable taking values in $\mathcal{X}$ and suppose that $Y = H(X)$ for some $H: \mathcal{X} \mapsto \mathcal{Y}$.  Further suppose that for some collection of functions $\mathcal{F}$ $\mathcal{X} \mapsto \mathcal{X}$,
 \begin{align}
 \label{eq-invariance-assumption}
  f(X) \stackrel{d}{=} X \ \ \forall \ f \in \mathcal{F}.  
 \end{align}
Furthermore, let $\mathcal{G}$ be a collection of functions $\mathcal{Y} \mapsto \mathcal{Y}$ and suppose that for any $g \in \mathcal{G}$, there exists a corresponding $f \in \mathcal{F}$ such that,
\begin{align}
\label{eq-transformation-assumption}
g(H(X)) = H(f(X)) \ \  a.s.    
\end{align}
Then,
\begin{align*}
g(Y) \stackrel{d}{=} Y \ \ \forall \ g \in \mathcal{G}.   
\end{align*}
\end{proposition}

Recall Assumption \ref{assumption-exchangeability} which states that the array $V_{ij} = (Y_i,Y_j, X_i, X_j, A_{ij})$ is jointly exchangeable.     We now make an additional assumption on network covariates.    Let $\zeta(A,X_1,\ldots, X_{2n+1})$ denote the function that outputs  $(\hat{Z}_1, \ldots, \hat{Z}_{2n+1})$, and for a permutation $\sigma$  let $A^{(\sigma)}$ be a matrix such that $A_{ij}^{(\sigma)} = A_{\sigma(i)\sigma(j)}$ for $i,j \in \{1, \ldots 2n+1 \}$. 
\begin{assumption}
\label{assumption-permutation-invariance}
For any permutation $\sigma: [2n+1] \mapsto [2n+1] $, the network covariates $(\hat{Z}_1, \ldots, \hat{Z}_{2n+1})$ satisfy:
\begin{align*}
(\hat{Z}_{\sigma(1)}, \ldots, \hat{Z}_{\sigma(2n+1)}) = \zeta(A^{(\sigma)}, (X_{\sigma(1)}, \ldots X_{\sigma(2n+1)})) \ \ a.s.
\end{align*}
\end{assumption}
In words, Assumption \ref{assumption-permutation-invariance} states that permuting the labels of the nodes results in permuting the vector of network covariates accordingly.
These two assumptions and Proposition \ref{theorem-exchangeability-transformations} lead to our first main result.  
\begin{theorem}[Validity of Split Conformal Prediction for Network Regression]
\label{theorem-conformal-guarantee}
Under Assumptions \ref{assumption-exchangeability} and \ref{assumption-permutation-invariance}, $(Y_i, X_i, \hat{Z}_i)_{1 \leq i \leq 2n+1}$ is exchangeable.  Moreover, for the split conformal prediction procedure defined in Algorithm 1,
\begin{align}
\label{eq-split-conformal-validity}
P(Y_{2n+1} \in \widehat{C}(X_{2n+1}, \hat{Z}_{2n+1})) \geq 1-\alpha.  
\end{align}
\end{theorem}

\begin{remark}
Analogous to other settings, if $S_{n+1}, \ldots, S_{2n+1}$ are almost surely distinct, we have the stronger guarantee:
\begin{align*}
1 - \alpha \leq P(Y_{n+1} \in \widehat{C}(X_{n+1}, \hat{Z}_{n+1})) \leq 1 - \alpha + \frac{1}{n+1}.
\end{align*}
\end{remark}

\begin{remark}
Other procedures that require only that the data points are exchangeable such as the original conformal prediction procedure and the jackknife+\citep{10.1214/20-AOS1965} are also valid under Assumptions \ref{assumption-exchangeability} and \ref{assumption-permutation-invariance}.  
\end{remark}

\begin{remark}
For the adjacency spectral embedding defined in Example \ref{example-ase}, the permutation invariance property is algorithm-dependent since eigenvector solvers need not be permutation invariant. 
Nevertheless, we do not expect this to be an issue in practice since the conformal prediction procedure itself does not directly use the property \eqref{eq-transformation-assumption} and there always exists a permutation-invariant eigenvector solver that produces the same solution on $A$. 
\end{remark}

Implicit in our construction of the network summary statistics is that $X_{2n+1}$ and $\{A_{i(2n+1)}\}_{1\leq i \leq 2n+1}$ are available when the nonconformity scores $\{S_i\}_{n+1 \leq i \leq 2n+1}$ are computed. We exclude $(Y_1, \ldots, Y_{2n+1})$ from the construction of the network covariate since $Y_{2n+1}$ is unknown; including the response would break the symmetry between $\hat{Z}_i$ and $\hat{Z}_{2n+1}$ for $1 \leq i \leq 2n$. 

However, in certain situations, one may one want to compute $S_i$ only using first $2n$ nodes or include the response variable in the network statistic.  In these cases, there are a few possible approaches.  If one is simply interested in including $(Y_1, \ldots, Y_{2n+1})$, one may consider the original conformal prediction procedure at the cost of increased computation.  One may also consider split network statistics of the form:
\begin{align*}
\widetilde{Z}_k = h\left( (V_{ij})_{i,j \in \{1, \ldots n\} \cup \{k\}} \right)
\end{align*}
for some measurable $h$ and $k \in \{n+1, \ldots, 2n+1\}$.  For instance, the split analog for the statistic considered in Example \ref{example-neighbor-average} is given by:
\begin{align}
\label{eq:split-network-statistic}
\widetilde{Z}_k = \frac{1}{ \sum_{j=1}^n A_{kj}}\sum_{j=1}^n A_{kj}X_j.
\end{align}
So long as a condition analogous to Assumption \ref{assumption-permutation-invariance} applies to permutations on $\{n+1, \ldots 2n+1\}$, the property \eqref{eq-split-conformal-validity}  may be proved for this variant of conformal prediction with arguments analogous to those used to prove Theorem \ref{theorem-conformal-guarantee}.    

While the split network statistic proposal preserves finite sample validity, splitting in this manner also leads to higher variability, which may be most noticeable for sparse graphs with a few influential nodes.  Instead, one may also consider leveraging the stability of model and score function, which yields asymptotic validity under some mild additional conditions.  We state this result below. 
\begin{proposition}
\label{prop:stability}
Suppose that $S_{n+1}, \ldots, S_{2n+1}$ are nonconformity scores where for $n+1 \leq i \leq 2n$, $\hat{Z}_i$ is a function of $(V_{ij})_{1 \leq i,j \leq 2n}$  and $\hat{Z}_{2n+1}$ is a function of  $(V_{ij})_{1 \leq i,j \leq 2n},$ $(A_{(2n+1)j})_{1 \leq j \leq 2n}, X_{2n+1}$.   Suppose there exist i.i.d.\ random variables $\widetilde{S}_{n+1}, \ldots, \widetilde{S}_{2n+1}$ such that $\widetilde{S}_i$ is continuous with a bounded density function and $\widetilde{S}_i - S_i = o_P(1)$. Then,
\begin{align}
\label{eq-network-conformal-guarantee}
P(Y_{2n+1} \in \widehat{C}(X_{2n+1}, \hat{Z}_{2n+1})) = 1-\alpha + o(1).   
\end{align}
\begin{remark}
While independence is strictly speaking not necessary, it rules out strong dependence between observations.  Independence also often holds in Setting \ref{setting-independent-triples}. 
\end{remark}
\begin{remark}
Continuity is an innocuous assumption since one may want to add a small amount of continuous noise to the nonconformity scores to break ties anyway; see \citet{kuchibhotla2021exchangeability} on ``jittering" or \citet{vovk-algorithmic-world} on smoothed conformal prediction.   
\end{remark}
\end{proposition}

\subsection{Asymptotic Conditional Validity}
\label{sec:conditional-validity}
In this section, we restrict our attention to Setting \ref{setting-independent-triples} and establish conditions under which asymptotic validity holds, that is, 
\begin{align}
\label{eq:conditional-validity-network}
P(Y_{2n+1} \in \hat{C}(X_{2n+1}, \hat{Z}_{2n+1}) \ | \ X_{2n+1}, Z_{2n+1} ) = 1 -\alpha + o_P(1).    
\end{align}
Here $\hat{Z}_{2n+1}$ is a network statistic approximating some population quantity $Z_{2n+1}$ and $(Y_{i},X_{i},Z_{i})_{1 \leq i \leq 2n+1}$ are assumed to be i.i.d.\ and continuous.   Note that we condition on the population quantity, which is unobserved; yet, the property \eqref{eq:conditional-validity-network} ensures that coverage is comparable throughout the space.  

The case of discrete $\hat{Z}_i$ is obviously of interest, since one may, for example, be interested in guaranteeing similar coverage across (discrete) network communities.  The discrete case is in some sense an easier problem and we believe that approaches previously considered in the literature by \citet{pmlr-v25-vovk12}, \citet{https://doi.org/10.1111/rssb.12021}, and \citet{10.1093/imaiai/iaaa017} can be extended to this setting without much difficulty.    

For ease of exposition, we restrict our attention to distribution regression methods, which aim to estimate the conditional CDF by fitting a collection of regressions over a grid of values for $t$.  The most straightforward approach is regressing the binarized response $\mathbbm{1}(Y \leq t)$ against covariates $X$ and $Z$ for each $t$ in the grid.

Our overall strategy is based on leveraging the stability of the fitted regression model.  In essence, if the model is stable enough and $\hat{Z}_i$ is close to $Z_i$,  the problem can be approximated by the independent case for which the results of \citet{Chernozhukove2107794118} apply.   We state a general result below and later verify the conditions for kernel regression.  

We first introduce some additional notation.  Let $\hat{F}_{Y | X,Z}^{(r)}(y | x,z)$ denote the conditional CDF estimator in which the training data consists of  
\begin{align*}
(Y_1,X_1,Z_1), \ldots, (Y_r,X_r,Z_r), (Y_{r+1},X_{r+1},\hat{Z}_{r+1}), \ldots, (Y_{n},X_{n},\hat{Z}_{n}),
\end{align*}
where $r=0$ corresponds to the original dataset. Further, let $\tilde{F}_{Y | X,Z}(y | x,z) = \hat{F}_{Y | X,Z}^{(n)}(y | x,z) $ and $F_Y(y|x,z)$ denote the CDF of $Y \ | \ X,Z$.  We have the following result:
\begin{theorem}[Asymptotic Conditional Validity of Split Conformal Prediction]
\label{theorem-asymp-cond-conformal-pred}
Suppose that we are in Setting \ref{setting-independent-triples} with $(Y_i, X_i, Z_i)_{1 \leq i \leq 2n+1}$ iid and continuous and that $\hat{Z}_1, \ldots, \hat{Z}_{2n+1}$ satisfy Assumption \ref{assumption-permutation-invariance}.  Moreover, suppose that the following conditions hold:
\begin{enumerate}
    \item[(a)]Training example stability. 
    \begin{align*}
    \max_{1 \leq r \leq n}  \max_{1 \leq i \leq n+1} \left|\hat{F}_{Y |  X,Z}^{(r)}(Y_{n+i} \ | \ X_{n+i}, Z_{n+i}) - \hat{F}_{Y | X,Z}^{(r-1)}(Y_{n+i} \ | \  X_{n+i}, Z_{n+i}) \right| = o_P(1/n)  
    \end{align*}
    \item[(b)]Input stability.
    \begin{align*}
    \max_{1 \leq i \leq n+1} \left| \hat{F}_n(Y_{n+i} \ | \ X_{n+i}, \hat{Z}_{n+i}) - \hat{F}_n(Y_{n+i} \ | \ X_{n+i}, Z_{n+i}) \right| = o_P(1) 
    \end{align*}
    \item[(c)]Consistency of distributional regression.
    \begin{align*}
    \mathbb{E}\left|\tilde{F}_{Y|X,Z}(Y_{2n+1} \ | \ X_{2n+1}, Z_{2n+1} ) -  F_{Y|X,Z}(Y_{2n+1} \ | \ X_{2n+1}, Z_{2n+1})\right| = o(1).
    \end{align*}
\end{enumerate}
Then, for the confidence set $\hat{C}_n(X_{2n+1}, \hat{Z}_{2n+1})$ constructed using Algorithm 1 with $S_i =|1/2 - \hat{F}_n(Y_i \ | \ X_i, \hat{Z}_i)|$, the asymptotic conditional validity property (\ref{eq:conditional-validity-network}) holds.  Moreover, the unconditional property (\ref{eq-split-conformal-validity}) continues to hold even if properties (a)-(c) are violated. 
\end{theorem}
 
The training example stability condition is related to the Lindeberg interpolation, which uses a telescoping sum to reduce the problem to bounding the effect of perturbing one observation at a time.  Many notions of stability in the literature also consider the effect of perturbing one data point at a time; see for example, \citet{bosquet-elisseeff-stability-genrealization}.   With central limit theorems, one uses the interpolation to bound expectations  related to a class of test functions; in our setting, it is also possible to consider stability in expectation rather than stability of a maximum with high probability.  However, convergence in probability is a weaker notion of convergence and is therefore easier to verify.  For example, high probability bounds for the maximum approximation error associated with random dot product graph embeddings have been attained by \citet{rubindelanchy2020statistical}, but to our knowledge, sharp bounds for the expectation have not been established.  
While the training example stability condition deals with replacing the training points with the true latent positions, the input stability deals with replacing the evaluation point with the true latent position.  Again, one may alternatively consider expectation bounds. Finally, for the consistency condition, results of this form are available for various estimators.  

We now consider kernel regression as an estimator for the conditional CDF.  Kernel methods are a natural nonparametric approach to this problem since they have the advantage of producing a valid CDF as an estimate when the kernel is non-negative. For distribution regression, variants of kernel regression have been studied by \citet{doi:10.1080/01621459.1999.10473832},  \citet{10.2307/27639002}, and \citet{Hansen2004}, among others.  In what follows let $\mathbb{R}^+$ denote the non-negative reals, $H: \mathbb{R}^{+} \mapsto \mathbb{R}^{+}$ and let $K: \mathbb{R}^q \mapsto \mathbb{R}^{+}$ denote the composition $H \circ f$ for some $f: \mathbb{R}^q  \mapsto \mathbb{R}^{+}$.   

For simplicity, we consider an estimator of the form
\begin{align}
\label{eq:kernel-regression-cdf}
\hat{m}(y,x,z) = \sum_{i=1}^n  \frac{K\left(\frac{\|X_i-x \| + \| \hat{Z}_i - z\|}{h} \right) \mathbbm{1}(Y_i \leq y)}{\sum_{i=1}^n  K\left(\frac{\|X_i-x \| + \| \hat{Z}_i - z\|}{h}\right)}.   
\end{align}
The kernel above takes as input norms of the conventional covariate and network covariates separately, which may be particularly useful for random dot product embeddings. See Remark \ref{remark:RDPG-conditional-validity} for further discussion of asymptotic  validity conditional on the latent position of an RDPG model. 

\begin{theorem}[Asymptotic Conditional Validity for Kernel Regression]
\label{theorem-conditional-validity-kernel-regression}
Suppose that $(Y_1, X_1, Z_1), \ldots, (Y_{2n+1}, X_{2n+1}, Z_{2n+1})$ are i.i.d.\ triples such that $(X_1,Z_1) \in \mathbb{R}^q$ is supported on a hyperrectangle of the form $\mathcal{R} = \prod_{j=1}^q [a_j, b_j]$ and that the joint density $f_{XZ}(x,z)$ satisfies $0 <c \leq f_{X,Z}(x,z) \leq C < \infty$ for all $(x,z) \in \mathcal{R}$. Moreover, suppose that the kernel $K(\cdot)$ satisfies the following conditions: 
\begin{enumerate}
    \item[(a)] Decreasing with light tails:   $H(\cdot)$ is a decreasing function satisfying $t^q H(t) \rightarrow 0$. 
    \item[(b)] Bounded: $H(0) < \infty$. 
    \item[(c)] Compact kernel lower bound: 
    For some $k,r > 0$, $K(x,z) \geq k \cdot \mathbbm{1}( \|(x,z) \|_2 \leq r)$.
    \item[(d)] Lipschitz:  $|H(x)-H(y)| \leq L|x-y| \ \forall \ x,y \in \mathbb{R}^{+}$ for some $0 < L <\infty$.
\end{enumerate}
Furthermore, suppose that $h_n \rightarrow 0$, $\frac{nh_n^q}{|\log h_n|} \rightarrow \infty$, $\frac{|\log h_n|}{\log \log n } \rightarrow \infty$, and
\begin{align*}
\max_{1 \leq i \leq 2n+1} \frac{\|\hat{Z}_i - Z_i \|}{h_n^{q+1}} = o_P(1).   
\end{align*}
Then asymptotic conditional validity \eqref{eq:conditional-validity-network} holds for scores of the form $S_i = |1/2 - \hat{F}_n(Y_i \ | \ X_i, \hat{Z}_i)|$ constructed from the kernel regression estimate \eqref{eq:kernel-regression-cdf}.   
\end{theorem}

Note that we place no conditions on the smoothness of the conditional CDF or the graphon.  In the sparse graphon model, irregularity of the graphon can capture certain features of real-world graphs such as heavy-tailed degree distributions and hubs; therefore, we believe that having minimal assumptions on the graphon itself is particularly important.  We invoke results on universal consistency of kernel regression, e.g.,  \citet{10.1214/aos/1176344949,10.1214/aos/1176346815}, to verify consistency under these mild conditions. 

We do require that $(X,Z)$ is supported on a hyperrectangle and that the joint density $f_{XZ}(x,z)$ is bounded above and below. 
The bounded support and lower bound on the density rule out potential low density areas where kernel regression may not be stable; it remains to be seen whether this assumption can be relaxed.  The geometry of hyperrectangles is used to lower bound a smoothed density uniformly, but other compact sets may be considered for the support.  The upper bound on the density arises from invoking a result due to \citet{AIHPB_2002__38_6_907_0} to uniformly control the behavior of a kernel density estimate.    

Finally, our conditions on the kernel and the bandwidth are standard.  The network statistic condition implies that the bandwidth depends on the rate of the convergence of the network functional, which will depend on the sparsity level of the graph. 
\begin{remark}
\label{remark:RDPG-conditional-validity}
For RDPG embeddings, one typically only has high probability bounds for $ \max_{1 \leq i \leq 2n+1} \|Q_n \hat{Z}_i - Z_i \|$, where $Q_n$ is an unknown (random) orthogonal rotation.  However,  for the choice of kernel \eqref{eq:kernel-regression-cdf} the scores do not change when a rotation is applied to the estimated latent positions. Thus, asymptotic validity given a conditioning set that includes the latent position of the test point still holds if $\max_{1 \leq i \leq 2n+1} \|Q_n \hat{Z}_i - Z_i \|$ can be adequately controlled  and the other conditions in Theorem \ref{theorem-conditional-validity-kernel-regression} are satisfied.        
\end{remark}

\section{Experiments}
\label{sec:experiments}   
\subsection{Unconditional Validity}
\label{sec:sims-unconditional}
For our study of unconditional validity, we consider two data generating processes and compare conformal prediction to a naive Gaussian prediction interval.  Our proposed method is thus far the only one we are aware of in the literature to offer guarantees for network-assisted regression, and therefore has no true competitor. However, in Setting \ref{setting-independent-triples}, if the error term is Gaussian, our fitted model is correctly specified, and network covariates converge to some node-level counterpart, then the Gaussian model is ``approximately correct" and is a reasonable comparison. We consider two sample sizes, $n=1000$ and $n=3000$, with the first half of the data used to train the model and the second half used to compute nonconformity scores.  The last data point is reserved to construct a prediction interval. We set $\alpha = 0.1$ and assess coverage over 500 replications. The sparsity parameter $\rho_n$ ranges from $n^{-0.1}$ to $n^{-0.75}$.            
\subsubsection{Linear Model with Random Dot Product Graph Embeddings}
The first data generating process we consider is a linear model, 
\begin{align*}
Y = 3 + 2X + 10Z_{1} + 15Z_{2} - 17 Z_{3} + \epsilon,  
\end{align*}
where $\epsilon \sim N(0,1)$, $X = U Z_{1} + W $, $U \sim \mathrm{Uniform}[1,2], W \sim N(0,1)$, and $(Z_{1},Z_{2}, Z_{3})$ are the latent positions of a rank 3 random dot product graph model.  While the latent positions are identifiable only up to orthogonal rotation, each rotation results in a linear model and predictions from a linear model are invariant under rotation of the underlying positions.  
The random dot product graph under consideration corresponds to a truncated eigendecomposition of the graphon $\min(x,y)$, which has the eigenvalue-eigenfunction pairs:
\begin{align*}
\lambda_k = \left(\frac{2}{(2k-1) \pi}\right)^2, \ \ \  \phi_k(x) = \sin\left(\frac{(2k-1)\pi x}{2} \right).   
\end{align*}
See \citet{xu2018rates} for a derivation of these spectral properties.  We assume that the observed adjacency matrix has mean matrix $\nu_n \mathbf{Z} \mathbf{Z}^T$, where $\mathbf{Z}$ is the $n \times 3$ matrix with the ith row corresponding to the latent position for node $i$ and $\nu_n \in \{n^{-0.1}, n^{-0.25}, n^{-0.5}, n^{-0.75}\}$.

We fit a linear regression model $\hat{\mu}(\cdot)$ with covariate $X$, and network covariates $\hat{Z}_1, \hat{Z}_2, \hat{Z}_3$ corresponding to a rank 3 adjacency spectral embedding.  We construct conformal prediction intervals using the nonconformity score $s(x,y,z_1,z_2,z_3) = |y - \hat{\mu}(x,y,z_1,z_2,z_3)|$. 

\begin{table}[h]
\centering
\renewcommand*{\arraystretch}{1.2}
\scalebox{0.8}{
\begin{tabular}{cccccc}
                 & \multicolumn{2}{c}{Conformal Prediction}                   &                       & \multicolumn{2}{c}{Parametric Normal}                      \\ 
\cline{1-3} \cline{5-6} 
\multicolumn{1}{ | c ||}{$\nu_n$}                     & \multicolumn{1}{c|}{Coverage} & \multicolumn{1}{c|}{Width} & \multicolumn{1}{c|}{} & \multicolumn{1}{c|}{Coverage} & \multicolumn{1}{c|}{Width} \\ \cline{1-3} \cline{5-6} 
\multicolumn{1}{|c||}{$n^{-0.1}$}  & \multicolumn{1}{c|}{0.88}     & \multicolumn{1}{c|}{8.61}  & \multicolumn{1}{c|}{} & \multicolumn{1}{c|}{0.89}     & \multicolumn{1}{c|}{8.80}  \\ \cline{1-3} \cline{5-6} 
\multicolumn{1}{|c||}{$n^{-0.25}$} & \multicolumn{1}{c|}{0.88}     & \multicolumn{1}{c|}{8.84}  & \multicolumn{1}{c|}{} & \multicolumn{1}{c|}{0.91}     & \multicolumn{1}{c|}{9.19}  \\ \cline{1-3} \cline{5-6} 
\multicolumn{1}{|c||}{$n^{-0.33}$} & \multicolumn{1}{c|}{0.89}     & \multicolumn{1}{c|}{8.89}  & \multicolumn{1}{c|}{} & \multicolumn{1}{c|}{0.92}     & \multicolumn{1}{c|}{9.35}  \\ \cline{1-3} \cline{5-6} 
\multicolumn{1}{|c||}{$n^{-0.75}$} & \multicolumn{1}{c|}{0.91}     & \multicolumn{1}{c|}{8.91}  & \multicolumn{1}{c|}{} & \multicolumn{1}{c|}{0.91}     & \multicolumn{1}{c|}{9.47}  \\ \cline{1-3} \cline{5-6} 
\end{tabular}}
\caption{Coverage and width of prediction intervals for the linear model with random dot product graph embeddings, $n=1000$.}
\label{table-rdpg-linear}
\end{table}

Results for $n=1000$ Are summarized in Table \ref{table-rdpg-linear}.  Results for $n=3000$ are similar, with very slight improvements in the average width for both methods, and are omitted for lack of space.     
As expected, the conformal prediction intervals have coverage close to the nominal level of $0.9$. For this particular example, the parametric normal prediction intervals also have coverage around $0.9$, since the fitted model is correct, but are wider on average.  If the true latent positions were known, the parametric normal intervals would be optimal; however, the estimation error in latent positions inflates the variance and leads to wider intervals.   Since higher sparsity leads to larger estimation error, it is not surprising that the average width increases faster for parametric normal intervals compared to the conformal prediction intervals as the network becomes more sparse.

\subsubsection{Spatial Autoregressive Model}
\begin{figure}[h]
\centering
\includegraphics[scale=0.4]{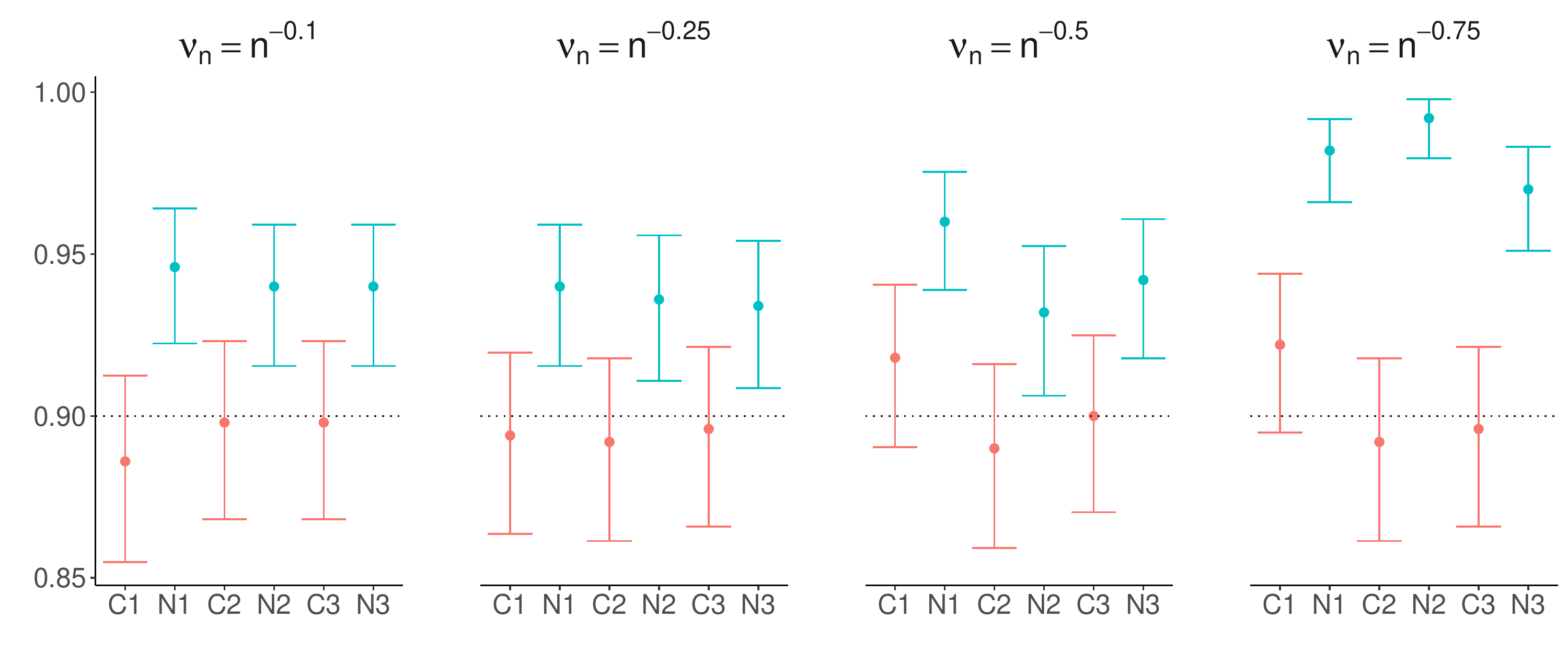}
\caption{Coverage of prediction intervals for the spatial autoregressive model with $n=3000$, where ``C'' denotes conformal prediction, ``N'' denotes parametric normal, and the number denotes the model number.  The color red corresponds to conformal prediction whereas the color blue corresponds to parametric normal. The displayed confidence intervals are exact binomial intervals for the coverage probability.  }
\label{fig:spat-auto-coverage}
\end{figure}

Here we consider a variant of the data generating process considered in Setting \ref{setting-neighborandnode}.  Recall that $\widetilde{Y}_i$ and $\widetilde{X}_i$ represent neighborhood averages of $Y$ and $X$, respectively, for node $i$.  Now, suppose that $Y_1, \ldots, Y_{2n+1}$ are generated by the following model:
\begin{align*}
Y_i =  4X_{1i} +5 X_{2i} + 0.7 \ \widetilde{Y}_i +2\widetilde{X}_{1i} + 3\widetilde{X}_{2i} + \epsilon_i,
\end{align*}
where $\epsilon_i \sim N(0,1)$ and $(X_{1i}, X_{2i}, Z_{i}) \sim N( \mu, \Sigma)$ with:
\begin{align*}
\mu = (1,3,0) , \ \
\Sigma = \begin{pmatrix}
1 & 0.6 & 0.3 \\ 
0.6 & 4 & \text{-} 0.4 \\ 
0.3 & \text{-} 0.4 & 1
\end{pmatrix}.
\end{align*}
Solving the above system of equations to find the implied values of $Y$ is straightforward; see, for example, Section 5.1 of \citet{10.1214/18-AOAS1205}.  For both sample sizes, we assume a population size of $N=3000$. We generate $A_{ij}$ from a Gaussian latent space model of the form:
\begin{align*}
A_{ij} \sim \mathrm{Bernoulli}(\nu_n \exp\{- (Z_i - Z_j)^2/4\}).   
\end{align*}
For both conformal prediction and the naive Gaussian prediction interval, we consider three different models, representing different levels of misspecification.  We first consider the practically important case where network information is unobserved, and fit the model 
$$\hat{\mu}_1(X_1, X_2) = \hat{\beta}_0 + \hat{\beta}_1 X_1 + \hat{\beta}_2 X_2 .$$ 
The second model incorporates network information, but does not account for endogeneity in $Y$: 
$$\hat{\mu}_2(X_1,X_2, \tilde{X}_1, \tilde{X}_2) = \hat{\beta}_0 + \hat{\beta}_1 X_1 + \hat{\beta}_2 X_2 + \hat{\beta}_3 \tilde{X}_1 + \hat{\beta}_4 \tilde{X}_2. $$  
Finally, $\hat{\mu}_3$ will be a correctly specified linear model. We exclude $Y_{2n+1}$ from the neighborhood average.  The regression model will be of the form:
$$\hat{\mu}_3(X_1,X_2, \tilde{X}_1, \tilde{X}_2, \tilde{Y}) = \hat{\beta}_0 + \hat{\beta}_1 X_1 + \hat{\beta}_2 X_2 + \hat{\beta}_3 \tilde{X}_1 + \hat{\beta}_4 \tilde{X}_2 + \hat{\beta}_5 \tilde{Y}. $$  
For the spatial autoregressive model, we present results for coverage in Figure \ref{fig:spat-auto-coverage}
and width in Table \ref{table:spat-auto-width}.    
As our theory predicts, the triples constructed from the spatial autoregressive model appear to be exchangeable; our conformal prediction intervals have coverage near the nominal level.  On the other hand, the parametric normal intervals are unnecessarily conservative, staying well above the $0.9$ nominal coverage for all models and sparsity levels and giving wider intervals than conformal prediction in all settings.   From Table \ref{table:spat-auto-width}, we can infer that the variance inflation is much more severe for this data generating process, which strongly depends on the neighbor-weighted response.   For the correctly specified model, the width of the conformal prediction intervals does not change much even when the problem becomes more difficult with increased sparsity, while we face a steeper penalty for model misspecification when the sparsity level increases, with the intervals for the misspecified models becoming much wider.  

\begin{table}[h]
\centering
\renewcommand*{\arraystretch}{1.2}
\scalebox{0.8}{
\begin{tabular}{lccclccc}
\multicolumn{1}{c}{}                      & \multicolumn{3}{c}{Conformal Prediction}                                                   &                       & \multicolumn{3}{c}{Parametric Normal}                                                      \\ \cline{2-4} \cline{6-8} 
\multicolumn{1}{l|}{}                     & \multicolumn{1}{c|}{Model 1} & \multicolumn{1}{c|}{Model 2} & \multicolumn{1}{c|}{Model 3} & \multicolumn{1}{l|}{} & \multicolumn{1}{c|}{Model 1} & \multicolumn{1}{c|}{Model 2} & \multicolumn{1}{c|}{Model 3} \\ \cline{1-4} \cline{6-8} 
\multicolumn{1}{|l|}{$\nu_n = n^{-0.1}$}  & \multicolumn{1}{c|}{3.57}   & \multicolumn{1}{c|}{3.30}   & \multicolumn{1}{c|}{3.30}   & \multicolumn{1}{l|}{} & \multicolumn{1}{c|}{4.24}   & \multicolumn{1}{c|}{3.93}   & \multicolumn{1}{c|}{3.93}   \\ \cline{1-4} \cline{6-8} 
\multicolumn{1}{|l|}{$\nu_n = n^{-0.25}$} & \multicolumn{1}{c|}{4.23}   & \multicolumn{1}{c|}{3.30}   & \multicolumn{1}{c|}{3.30}   & \multicolumn{1}{l|}{} & \multicolumn{1}{c|}{5.03}   & \multicolumn{1}{c|}{3.94}   & \multicolumn{1}{c|}{3.94}   \\ \cline{1-4} \cline{6-8} 
\multicolumn{1}{|l|}{$\nu_n = n^{-0.5}$}  & \multicolumn{1}{c|}{8.32}   & \multicolumn{1}{c|}{3.45}   & \multicolumn{1}{c|}{3.37}   & \multicolumn{1}{l|}{} & \multicolumn{1}{c|}{9.91}  & \multicolumn{1}{c|}{4.11}   & \multicolumn{1}{c|}{4.01}   \\ \cline{1-4} \cline{6-8} 
\multicolumn{1}{|l|}{$\nu_n = n^{-0.75}$} & \multicolumn{1}{c|}{27.21}  & \multicolumn{1}{c|}{8.60}  & \multicolumn{1}{c|}{3.34}   & \multicolumn{1}{l|}{} & \multicolumn{1}{c|}{37.14}  & \multicolumn{1}{c|}{15.11}  & \multicolumn{1}{c|}{4.77}   \\ \cline{1-4} \cline{6-8} 
\end{tabular}}
\caption{Average width of prediction intervals for the spatial autoregressive model with $n=3000$.}
\label{table:spat-auto-width}
\end{table}

\subsection{Conditional Validity}
In this section, we consider a highly heteroscedastic, nonlinear process and compare the performance of conformal prediction using the conditional CDF (see Section \ref{sec:conditional-validity}) to conformal prediction with the absolute residual as a nonconformity score. For sample size and sparsity, we consider the same experimental settings implemented in Section \ref{sec:sims-unconditional}.  To compare performance of the methods across the covariate space, for each method we use smoothing splines to estimate the coverage probability as a function of the covariate.  We present the corresponding curves and assess how close the curves are to nominal level of $0.9$.     

For simplicity, we consider a data generating process that depends only on one network covariate. Let 
\begin{align*}
Y_i = 4+5 \sin(3 \pi Z_i) + \frac{\exp(15Z_i)}{250} \ \epsilon_i,
\end{align*}
where $\epsilon_i \sim N(0,1)$ and $Z_i$ corresponds to the expected degree function of the graphon $w(\xi_i,\xi_j) = |\xi_i-\xi_j|$, given by
\begin{align*}
Z_i = \mathbb{E}[w(\xi_i,\xi_j) \ | \ \xi_i] = \xi_i^2 - \xi_i + 1/2   
\end{align*}
for $\xi_i \sim \mathrm{Uniform}[0,1]$.  We consider Setting \ref{setting-independent-triples}, the independent triples + sparse graphon model.  

Since this network statistic is in principle unobservable, in fitted models, we instead use the degree as the covariate. For conformal prediction based on the estimated conditional CDF, we consider kernel regression with cross-validated bandwidth selection.  We also consider two conformal prediction intervals that use the absolute error of the regression function as the nonconformity score: one interval uses linear regression as the fitted model and the other uses smoothing splines with degrees of freedom selected via cross-validation.        

\vspace{-5mm}
\begin{figure}[h]
\centering
  \subfloat[$n=1000,  \nu_n=n^{-0.1}$]{\includegraphics[scale=0.8]{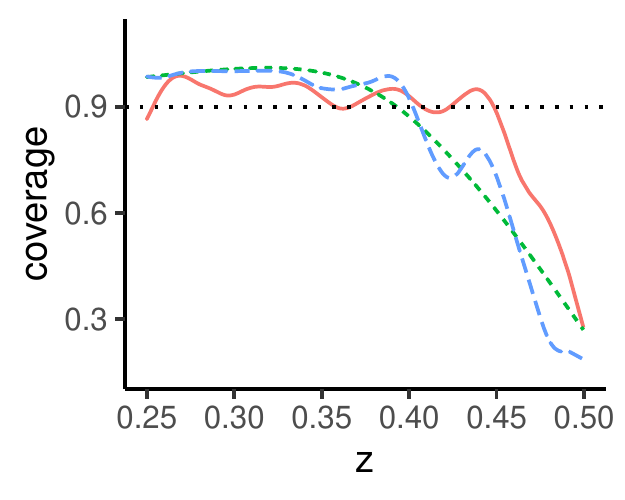}}
 \subfloat[$n=3000,  \nu_n=n^{-0.1}$]{\includegraphics[scale=0.8]{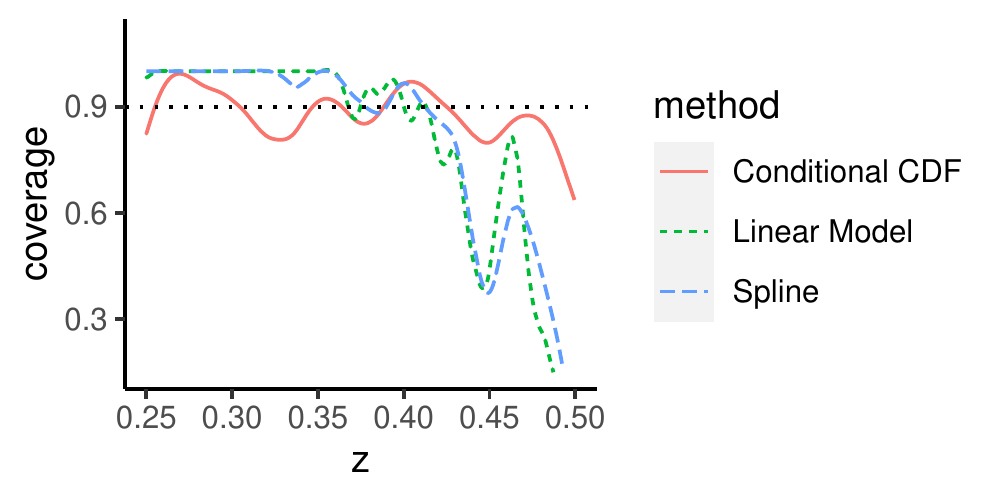}}
\caption{Comparison of conditional coverage for conformal prediction intervals.} 
\label{fig:cond1}
\end{figure}
\begin{figure}[h]
\centering
  \subfloat[$n=3000,  \nu_n=n^{-0.25}$]{\includegraphics[scale=0.8]{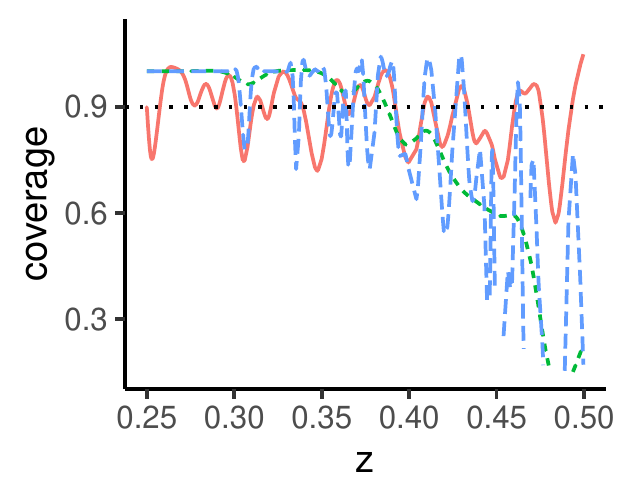}}
 \subfloat[$n=3000,  \nu_n=n^{-0.75}$]{\includegraphics[scale=0.8]{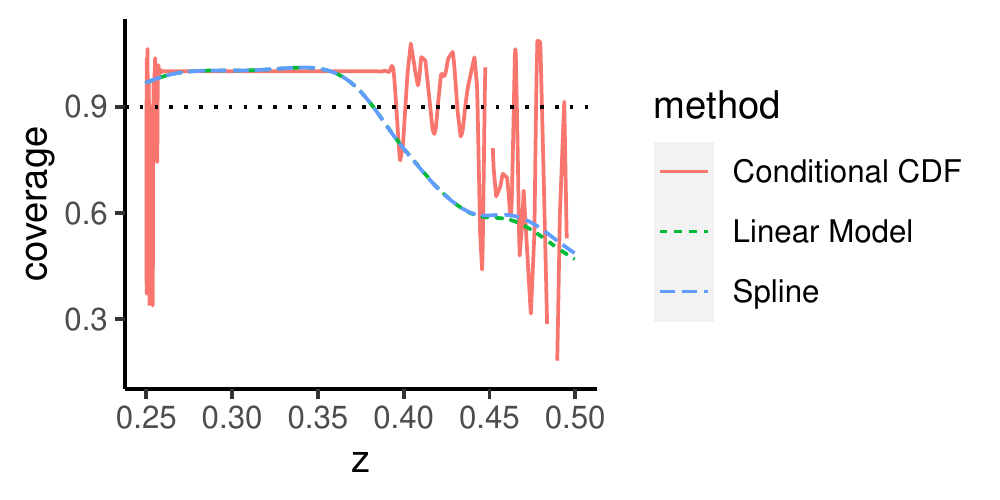}}
\caption{Conditional coverage of conformal prediction intervals as sparsity increases.}
\label{fig:cond2}
\end{figure}

Our simulation results for conditional validity are summarized in Figures \ref{fig:cond1} and \ref{fig:cond2}.  It is clear that conformal prediction intervals based on the standard regression nonconformity score do not achieve conditional validity.  In fact, these intervals tend to substantially overcover for regions with low variance and badly undercover for regions with high variance, with estimated coverage dipping to as low as 0.3 in some cases.  While we do not claim that our data generating process in this example is realistic,  it serves to highlight the fact that regression-based conformal prediction intervals may not be reliable under heteroscedasticity. 

In Figure \ref{fig:cond1}, the conditional CDF approach does not do much better than the other intervals for $n=1000$, but we see dramatic improvements at $n=3000$. As our theory predicts, for conditional validity, we need to estimate the conditional CDF well.  With network data, we have the additional complication that only noisy versions of the covariates are observed, which leads to slower convergence rates compared to i.i.d.\ data. 

In Figure \ref{fig:cond2}, we compare conformal prediction intervals as the sparsity level increases for $n=3000$.  For $\nu_n = n^{-0.25}$, the coverage of the interval based on the conditional CDF is slightly more erratic than what was observed in Figure \ref{fig:cond1} for $\nu_n = n^{-0.1}$ due to increased variance in the estimation of the covariate, but coverage for this approach is still much better for ``difficult'' regions of the space compared to the standard nonconformity score.  

For very sparse graphs with $\nu_n = n^{-0.75}$, we see that the conditional CDF estimate deteriorates further, to the point where coverage is not approximately equal throughout the space.  In this regime, smaller values of $Z_i$ become harder to distinguish from $0$ even though the random variable is bounded away from $0$, leading to very noisy estimates of the CDF in this range.  Although coverage may still be better than the other intervals for larger values of $z$, we still see more deterioration in performance compared to other sparsity levels.  Our simulation results here suggest that approximate conditional validity may be too ambitious of a target for very sparse graphs.  The situation is likely much worse when more covariates are included in the model due to the curse of dimensionality.      

\subsection{Example:  Classifying Machine Learning Papers}
In this section, we illustrate our split conformal prediction procedure on the Cora dataset \citep{cora-paper}, which consists of 2708 machine learning papers classified into 7 different categories.  We consider the task of predicting whether or not the paper category is ``Neural Networks'',  using both citations, which come in the form of a network, and conventional features extracted from a $2708 \times 1433$ document-word matrix, where the columns contain counts of the most frequently appearing words; this classification problem was previously studied by \citet{JMLR:v21:17-470}.   
We consider several different models and evaluate coverage on a holdout set of 500 observations, using the remaining 2208 observations for training and validation.
The predictive models that we evaluate include random forests and logistic regression with various covariates.   


For all models, we include top 20 principal components of the document-word matrix, which we call covariate set 1.  
Covariate set 2  includes node degrees and random dot product graph embeddings (see Example \ref{example-ase})  with $d=3$.  
Covariate set 3 contains estimated parameters from the model introduced in \citet{JMLR:v21:17-470}, given by
\begin{align*}
\text{logit}(P_{ij}) = \alpha_i + \alpha_j + z_i^T z_j,   
\end{align*}
where $\alpha$ is a degree heterogeneity parameter and $z_i$ is the embedding. To fit the model, we consider a random initialization and implement the projected stochastic gradient descent algorithm proposed by the authors, again assuming $d=3$ for the embedding.
For covariate set 4, we include degrees and neighborhood averages of $Y$. 
For the latter, we consider the split network statistic proposed in \eqref{eq:split-network-statistic}.  We also consider certain unions of the covariate sets; for example, we consider models with the covariate set $\{1,2,4\}$ and $\{1,3,4\}$. Since the degree heterogenity parameters $\alpha_i$ and $\alpha_j$ for the logit model are closely related to degrees, we exclude degrees for the the covariate set $\{1,3,4\}$.       

To construct conformal prediction sets, we take estimated conditional class probabilities from our fitted model and use the nonconformity score studied in \citet{NEURIPS2020_244edd7e}. We again consider $\alpha =0.1$, which we found leads to nontrivial prediction sets for all three models. 
\begin{table}[h]
\centering
  \renewcommand*{\arraystretch}{1.1}
 \scalebox{0.8}{
\begin{tabular}{cccclccc}
\multicolumn{1}{l}{}               & \multicolumn{3}{c}{Logistic Regression}                                                         &                       & \multicolumn{3}{c}{Random Forest}                                                               \\ \cline{1-4} \cline{6-8} 
\multicolumn{1}{|l||}{Covariates}   & \multicolumn{1}{l|}{Error rate} & \multicolumn{1}{l|}{Coverage} & \multicolumn{1}{l|}{\ Set size \ } & \multicolumn{1}{l|}{} & \multicolumn{1}{l|}{Error rate} & \multicolumn{1}{l|}{Coverage} & \multicolumn{1}{l|}{\ Set size \ } \\ \cline{1-4} \cline{6-8} 
\multicolumn{1}{|c||}{$\{1 \}$}     & \multicolumn{1}{c|}{0.16}       & \multicolumn{1}{c|}{0.93}     & \multicolumn{1}{c|}{1.24}     & \multicolumn{1}{l|}{} & \multicolumn{1}{c|}{0.19}       & \multicolumn{1}{c|}{0.90}     & \multicolumn{1}{c|}{1.21}     \\ \cline{1-4} \cline{6-8} 
\multicolumn{1}{|c||}{$\{1,2 \}$}   & \multicolumn{1}{c|}{0.16}       & \multicolumn{1}{c|}{0.93}     & \multicolumn{1}{c|}{1.24}     & \multicolumn{1}{l|}{} & \multicolumn{1}{c|}{0.19}       & \multicolumn{1}{c|}{0.91}     & \multicolumn{1}{c|}{1.22}     \\ \cline{1-4} \cline{6-8} 
\multicolumn{1}{|c||}{$\{1,3 \}$}   & \multicolumn{1}{c|}{0.15}       & \multicolumn{1}{c|}{0.93}     & \multicolumn{1}{c|}{1.21}     & \multicolumn{1}{l|}{} & \multicolumn{1}{c|}{0.19}       & \multicolumn{1}{c|}{0.90}     & \multicolumn{1}{c|}{1.24}     \\ \cline{1-4} \cline{6-8} 
\multicolumn{1}{|c||}{$\{1,4 \}$}   & \multicolumn{1}{c|}{0.14}       & \multicolumn{1}{c|}{0.91}     & \multicolumn{1}{c|}{1.14}     & \multicolumn{1}{l|}{} & \multicolumn{1}{c|}{0.14}       & \multicolumn{1}{c|}{0.89}     & \multicolumn{1}{c|}{1.04}     \\ \cline{1-4} \cline{6-8} 
\multicolumn{1}{|c||}{$\{1,2,4 \}$} & \multicolumn{1}{c|}{0.13}       & \multicolumn{1}{c|}{0.92}     & \multicolumn{1}{c|}{1.14}     & \multicolumn{1}{l|}{} & \multicolumn{1}{c|}{0.13}       & \multicolumn{1}{c|}{0.89}     & \multicolumn{1}{c|}{1.07}     \\ \cline{1-4} \cline{6-8} 
\multicolumn{1}{|c||}{$\{1,3,4 \}$} & \multicolumn{1}{c|}{0.15}       & \multicolumn{1}{c|}{0.93}     & \multicolumn{1}{c|}{1.20}     & \multicolumn{1}{l|}{} & \multicolumn{1}{c|}{0.14}       & \multicolumn{1}{c|}{0.89}     & \multicolumn{1}{c|}{1.04}     \\ \cline{1-4} \cline{6-8} 
\end{tabular}}
\caption{Results for the Cora dataset. Reported error (misclassification) rate, coverage, and set size are averages over 500 points in the test set.}
\label{table:cora-data}
\end{table}

We present results for our data analysis in Table \ref{table:cora-data}.  It appears that vertex exchangeability is a reasonable assumption for this dataset, as the coverage on the holdout set is consistent with our theory. Note that for the logit embeddings, we have no guarantees that the estimated model corresponds to a global optima; however, as our theory suggests, this does not affect the coverage properties of conformal prediction.  In this dataset, it appears that, for the most part, the inclusion of network covariates yields only a modest improvement in the misclassification rate.  However, the improvement in the average width (average set size) of the prediction sets is more noticeable.  Unsurprisingly, from comparing widths of prediction intervals,    
it appears that the neighbor-weighted response is an informative covariate.  With our nonconformity score, the accuracy of the estimated conditional probability also matters; while the inclusion of network covariates may not drastically impact the classification decision, it appears that it does improve the fitted model, leading to smaller average set sizes.   
It is also not surprising that random forests provided a better fit than logistic regression and produced prediction sets with smaller width overall.    

\section{Discussion}
\label{sec:discussion}
The main contribution of this work is leveraging a form of vertex exchangeability to extend the validity of conformal prediction to network-assisted prediction.  The generality of our approach suggests that conformal prediction may be used for a wide range of problems making use of network data, for example, in link prediction, which we are currently investigating.   While conformal prediction provides a useful quantification of uncertainty under mild assumptions, it does not offer natural confidence intervals for underlying parameters, particularly those related to expectations. Further developing inferential methods and theory for network-assisted regression is an important future direction, and the exchangeability results may come in useful for that as well.        

\section*{Acknowledgements}
The authors would like to thank Weijing Tang for sharing code for fitting the logit latent space model.   EL's research is supported by NSF grants 1916222, 2052918,  2210439. JZ's research is supported by NSF grants 1821243, 2123777, 2210439.  

\bibliographystyle{apalike}
\bibliography{mybib.bib}
\newpage

\section*{Appendix}

\textit{Proof of Theorem \ref{theorem-exchangeability-transformations}}.
For a given $g \in \mathcal{G}$ and any measurable set $A \subseteq \mathcal{Y}$, we can choose $f \in \mathcal{F}$ so that:
\begin{align*}
 P(g(Y) \in A) &= P( H(f(X)) \in A) && \text{(Using Assumption \ref{eq-transformation-assumption})}
 \\ &= P( f(X) \in H^{-1}[A])
 \\ &= P( X \in H^{-1}[A]) && \text{(Using Assumption \ref{eq-invariance-assumption})}
 \\ &= P( Y \in A).
\end{align*}
Since the choice of $g \in \mathcal{G}$ was arbitrary, the result follows. \qed 
\bigskip

\noindent \textit{Proof of Theorem \ref{theorem-conformal-guarantee}}. Let $\mathcal{Y} = \mathbb{Y}^{2n+1}$ where $\mathbb{Y} \subseteq \mathbb{R} \times \mathbb{R}^d \times \mathbb{R}^p$ and $\mathcal{X} = \mathbb{V}^{(2n+1) \times (2n+1)}$, where $V_{ij} \in \mathbb{V}$.  Taking $\mathcal{G}$ to be the class of permutations, by Assumption \ref{assumption-permutation-invariance}, for each $g \in \mathcal{G}$, we can choose a $f \in \mathcal{F}$ corresponding to the same permutation acting on the rows and columns of $V$. Since for each  permutation $f(V) \stackrel{d}{=}V$ by Assumption \ref{assumption-exchangeability}, exchangeability of the triples $(Y_{i},X_{i}, \hat{Z}_{i})_{1 \leq i \leq 2n+1}$ follows from Proposition \ref{theorem-exchangeability-transformations}. 

Now, for the second claim, observe that any permutation $\sigma$ of $(S_i)_{n+1 \leq i \leq 2n+1},$ corresponds to:
\begin{align*}
   H((Y_{\sigma(i)},X_{\sigma(i)},\hat{Z}_{\sigma(i)})_{1 \leq i \leq 2n+1}) =  s(Y_{\sigma(i)},X_{\sigma(i)},\hat{Z}_{\sigma(i)}; \mathcal{D}_1)_{(n+1) \leq i \leq (2n+1)},
\end{align*}  
where $\sigma$ keeps $[n]$ fixed. Thus, the nonconformity scores are exchangeable, and (\ref{eq-network-conformal-guarantee}) follows from properties of quantiles of exchangeable random variables (see, for example, Lemma 1 of \citet{NEURIPS2019_8fb21ee7}).
    \qed \\  

\noindent \textit{Proof of Proposition \ref{prop-setting-exchangeable}}.
For Setting \ref{setting-independent-triples}, by Proposition \ref{theorem-exchangeability-transformations}, it can be readily seen that $(Y_{i}, Y_{j}, \xi_i, \xi_j, \eta_{ij})_{1 \leq i,j \leq 2n+1}$ is  jointly exchangeable.  One may then construct $A_{ij} = \mathbbm{1}(\eta_{ij} \leq \rho_n w(\xi_i,\xi_j))$ and invoke Proposition  \ref{theorem-exchangeability-transformations} once more to verify joint exchangeability of $(V)_{1 \leq i,j \leq 2n+1}$.   

For Setting \ref{setting-neighborandnode}, we first establish exchangeability of the vector $U = (U_1, \ldots, U_N)$, where:
\begin{align*}
U_i = (X_i, \xi_i, \epsilon_i,\widetilde{D}_i^{(1)}, \ldots, \widetilde{D}_i^{(N-1)}, \widetilde{X}_i^{(1)}, \ldots, \widetilde{X}_i^{(N-1)},\eta_{i1} ,\ldots, \eta_{iN}).
\end{align*}
Note that shortest path lengths between two given nodes are invariant to joint permutations of the adjacency matrix; therefore Assumption \ref{assumption-permutation-invariance} is satisfied for the vector $(U_1, \ldots, U_N)$ and since $(X_i,X_j, \xi_i, \xi_j,\epsilon_i, \epsilon_j, \eta_{ij})_{1\leq i,j \leq N}$ is jointly exchangeable, $(U_1, \ldots, U_N)$ is exchangeable by Proposition \ref{theorem-exchangeability-transformations}. 
Now, we will define $G(\cdot)$ to be the function acting on $U$ that returns the vector $(y_1,U_1) ,\ldots (y_{N}, U_{N})$, where  $(y_1, \ldots y_N)$ is solution to the system of equations:
\begin{align*}
y_1 &= f\left( \frac{1}{\sum_{j\neq 1} \beta_{1j} } \sum_{j \neq 1}\beta_{1j}y_j \ ; U_1   \right)  \\ 
 \vdots & \\ 
y_N &= f\left( \frac{1}{\sum_{j\neq N} \beta_{Nj} } \sum_{j \neq N} \beta_{Nj}y_j \ ; U_N   \right).  
\end{align*}
When $U_1, \ldots, U_N$ are permuted, the system of equations is given by:
\begin{align*}
y_{1}' &= f\left( \frac{1}{\sum_{j\neq \sigma(1)} \beta_{\sigma(1)\sigma(j)} } \sum_{j \neq 1} \beta_{\sigma(1)\sigma(j)}y_{j}' \ ; U_{\sigma(1)}   \right)  \\ 
 \vdots & \\ 
y_{N}' &= f\left( \frac{1}{\sum_{j\neq \sigma(N)} \beta_{\sigma(N)\sigma(j)} } \sum_{j \neq N} \beta_{\sigma(N)\sigma(j)}y_{j}' \ ; U_{\sigma(N)}   \right). 
\end{align*}
This system has the solution $y_i' = y_{\sigma(i)}$ for $1 \leq i \leq N$ since
\begin{align*}
y_{\sigma(1)} &= f\left( \frac{1}{\sum_{j\neq \sigma(1)} \beta_{\sigma(1)\sigma(j)} } \sum_{j \neq 1} \beta_{\sigma(1)\sigma(j)}y_{\sigma(j)} \ ; U_{\sigma(1)}   \right)  \\ 
 \vdots & \\ 
y_{\sigma(N)} &= f\left( \frac{1}{\sum_{j\neq \sigma(N)} \beta_{\sigma(N)\sigma(j)} } \sum_{j \neq N} \beta_{\sigma(N)\sigma(j)}y_{\sigma(j)} \ ; U_{\sigma(N)}   \right) 
\end{align*}
is equivalent the original system of equations. Therefore, for each valid $(y_1,U_1(\omega) ,\ldots (y_{N}, U_{N}(\omega))$, and each permutation $\sigma$, we have that:
\begin{align*}
(y_{\sigma(1)},U_{\sigma(1)}(\omega) ) ,\ldots, (y_{\sigma(N)}, U_{\sigma(N)}(\omega)) = G(U_{\sigma(1)}(\omega),\ldots,U_{\sigma(N)}(\omega)).
\end{align*}   
Proposition \ref{theorem-exchangeability-transformations} now implies exchangeability of the pairs of interest since the theorem allows restriction to valid pairs.

Next, consider the mapping $F$ which constructs the array $(V_{ij})_{1 \leq i,j \leq 2n+1}$ from $(Y_1,U_1) ,\ldots, (Y_{N},U_{N})$.  For each joint permutation of the array $\sigma:[2n+1] \mapsto [2n+1]$, we see that this corresponds to $F((Y_{\sigma(1)},U_{\sigma(1)}), \ldots,  (Y_{\sigma(2n+1)} U_{\sigma(2n+1)}))$ and we may invoke Proposition \ref{theorem-exchangeability-transformations} once more, completing the proof. \qed  
\bigskip
\\ \textit{Proof of Proposition \ref{prop:stability}.}
Let $K = \sup_{x \in \mathbb{R}} f_{s_i}(x)$.  For any $\delta >0$, we have:
\begin{align*}
& \frac{1}{n+1}\sum_{i=n+1}^{2n} \mathbbm{1}(S_{2n+1} \geq S_i) \\ 
\leq & \ \frac{1}{n+1} \sum_{i=n+1}^{2n} \mathbbm{1}\left(\widetilde{S}_{2n+1} \geq \widetilde{S}_i - 2 \delta/K \right) + \mathbbm{1}\left( |\widetilde{S}_{2n+1} - S_{2n+1}| > \delta/K \right)  
\\ + & \  \frac{1}{n+1} \sum_{i=n+1}^{2n} \mathbbm{1}\left( |\widetilde{S}_i - S_i| > \delta/K \right) \\    
= & \ I + II + III.
\end{align*}
Since $\tilde{S}_i - S_i = o_P(1)$, for $n$ large enough, $P(II > \delta) \leq \delta$ and $P(III > \delta) \leq \delta$.

For $I$, observe that we may further upper bound with:
\begin{align*}
I &\leq \frac{1}{n+1} \sum_{i=n+1}^{2n} \mathbbm{1}\left(\widetilde{S}_{2n+1} \geq \widetilde{S}_i \right) + \frac{1}{n+1} \sum_{i=n+1}^{2n} \mathbbm{1}\left( \widetilde{S}_{2n+1} \leq \widetilde{S}_i \leq \widetilde{S}_{2n+1} + 2 \delta/K   \right)   
\\ &= I_A + I_B.
\end{align*}
Observe that:
\begin{align*}
\mathbb{E}{I_B} &=  P(\{\widetilde{S}_{2n+1} \leq \widetilde{S}_i \leq \widetilde{S}_{2n+1} + 2 \delta/K \}) \leq 2 \delta.
\end{align*}
Moreover, $|I_B- E(I_B)| = o_P(1)$.  Now, define the event:
\begin{align*}
B = \{II \leq \delta\} \cap \{III \leq \delta\} \cap \{|I_B- \mathbb{E}[I_B]| \leq \delta \}.
\end{align*}
Now, by inclusion-exclusion principle, for $n$ large enough:
\begin{align*}
P(Y_{2n+1} \in \widehat{C}_n(X_{2n+1}, \hat{Z}_{2n+1}))
\geq & \ P\left(I_A + I_B + II  + III \leq  \frac{\lceil(n+1)(1-\alpha)\rceil}{n+1} \ \cap \ B  \right) \\
\geq & \  P\left(I_A \leq  \frac{\lceil(n+1)(1-\alpha)\rceil}{n+1} - 5 \delta \right)  - P(II > \delta) \\ 
& - P(III > \delta)- P(|I_B - E(I_B)| > \delta ) 
\\ \geq &  \ P(I_A \leq 1 - \alpha - 5 \delta) - 3 \delta
\\ \geq & \  1- \alpha - 9 \delta,
\end{align*}
where in the last line we used the fact that when $\widetilde{S}_i$ are continuous, they are almost surely distinct and the rank is uniformly distributed on $\{1, \ldots, n+1 \}$.  The upper bound is analogous. \qed
\bigskip

\noindent \textit{Proof of Theorem \ref{theorem-asymp-cond-conformal-pred}}. 
We verify the conditions in Proposition 1 of \citet{Chernozhukove2107794118}. To check condition 1, observe that, conditions (a) - (c) imply that:

\begin{align*}
\frac{1}{n+1} \sum_{i=n+1}^{2n+1} \left| S_i - \widetilde{S}_i \right| &= \frac{1}{n+1} \sum_{i=n+1}^{2n+1} \left| \hat{F}_{Y|X,Z}(Y_i|X_i, \hat{Z}_i) - F_{Y|X,Z}(Y_i \ | \ X_i, Z_i) \right|
\\  \leq \ & \   \frac{1}{n+1} \sum_{i=n+1}^{2n+1} \left|\hat{F}_{Y|X,Z}(Y_i \ | \ X_i, \hat{Z}_i) -  \tilde{F}_{Y|X,Z}(Y_i \ | \ X_i, Z_i) \right| 
\\  & +  \frac{1}{n+1} \sum_{i=n+1}^{2n+1}\left|\tilde{F}_{Y|X,Z}(Y_i \ | \ X_i, Z_i) -  F_{Y|X,Z}(Y_i \ | \ X_i, Z_i)\right|
\\ \leq \ & \  \max_{1 \leq i \leq n+1} \left| \hat{F}_{Y |  X,Z}(Y_{n+i} \ | \ X_{n+i}, \hat{Z}_{n+i}) - \hat{F}_{Y |  X,Z}(Y_{n+i} \ | \ X_{n+i}, Z_{n+i})   \right| 
\\  & +  n \cdot \max_{1 \leq r \leq n}  \max_{1 \leq i \leq n+1} \left|\hat{F}_{Y |  X,Z}^{(r)}(Y_{n+i} \ | \ X_{n+i}, Z_{n+i}) - \hat{F}_{Y | X,Z}^{(r-1)}(Y_{n+i} \ | \  X_{n+i}, Z_{n+i}) \right|  
\\  & + \frac{1}{n+1} \sum_{i=n+1}^{2n+1} \left| \tilde{F}_{Y |  X,Z}(Y_{n+i} \ | \ X_{n+i}, Z_{n+i}) - F_{Y |  X,Z}(Y_{n+i} \ | \ X_{n+i}, Z_{n+i})   \right|
\\ = \ & \ o_P(1)
\end{align*}

Moreover, since $(Y_i,X_i,Z_i)$ are IID, condition 2 holds due to the DKW inequality, for example. For condition 3, observe that the random variable $|1/2 - F_{Y|X,Z}(Y_{2n+1} | X_{2n+1}, Z_{2n+1})| \sim \mathrm{Uniform}[0,1/2]$, which has a Lipschitz CDF.  Finally Assumption (c) guarantees Proposition 1 holds with $F^* = F$ and thus Theorem 3 of \citet{Chernozhukove2107794118} applies.  \qed
\bigskip

\noindent \textit{Proof of Theorem \ref{theorem-conditional-validity-kernel-regression}}. 
Let $\mathcal{C} = \mathbb{R} \times \mathcal{R}$.  To establish training example stability, we will show a stronger statement; namely the supremum over $\mathcal{C}$ is sufficently small.  We have that:
\begin{align*}
& \max_{1 \leq i \leq n} \sup_{(y,x,z) \in \mathcal{C}} \left|\hat{F}_{Y |  X,Z}^{(i)}(y \ | \ x, z) - \hat{F}_{Y | X,Z}^{(i-1)}(y \ | \ x, z) \right| \\ 
= & \max_{1 \leq i \leq n} \sup_{(y,x,z) \in \mathcal{C}} \left| \sum_{j=1}^n\frac{a_j^{(i)} \mathbbm{1}(Y_j \leq y) }{ \sum_{k=1}^na_k^{(i)}} - \sum_{j=1}^n\frac{a_j^{(i-1)} \mathbbm{1}(Y_j \leq y) }{ \sum_{k=1}^na_k^{(i-1)}} \right|,
\end{align*}
where $a_j^{(i)} = K(\frac{\|X_j -x\| + \|Z_j - z  \| }{h})$ for $j =1,\ldots i$ and $a_j^{(i)} = K(\frac{\|X_i -x\| + \|\hat{Z}_j - z  \| }{h})$ for $j=i+1, \ldots n$.  

Now, observe that:
\begin{align*}
\left| \ \frac{\partial}{\partial a_j^{(i)}} \left[\frac{ \sum_{k=1}^na_k^{(i)} \mathbbm{1}(Y_k \leq y) }{ \sum_{k=1}^na_k^{(i)}} \right] \  \right| &= \left|\frac{ (\sum_{k=1}^na_k^{(i)}) \cdot \mathbbm{1}(Y_j \leq y) - (\sum_{k=1}^na_k^{(i)} \mathbbm{1}(Y_k \leq y) )  }{ (\sum_{k=1}^na_k^{(i)})^2 } \right|
\\ & \leq \frac{2}{\sum_{k=1}^n a_k^{(i)}}.
\end{align*}
Therefore, by a Taylor expansion, 
\begin{align*}
& n \cdot \max_{1 \leq i \leq n} \ \sup_{(y,x,z) \in \mathcal{C} } \left| \sum_{j=1}^n\frac{a_j^{(i)} \mathbbm{1}(Y_j \leq y) }{ \sum_{k=1}^na_k^{(i)}} - \sum_{j=1}^n\frac{a_j^{(i-1)} \mathbbm{1}(Y_j \leq y) }{ \sum_{k=1}^na_k^{(i-1)} } \right| \\ 
\leq \ &  \ \max_{1 \leq i \leq n} \ \sup_{(x,z) \in \mathcal{R} } \left\{ \ \left|\frac{2}{\frac{1}{nh^q}\sum_{k  \neq i} a_k^{(i)} + \frac{1}{nh^q}\zeta_i^{(x,z)} } \right| \cdot \frac{|a_i^{(i)} - a_i^{(i-1)}|}{h^q} \  \right\}
\\ \leq \ & \  \left\{ \ \max_{1 \leq i \leq n} \ \sup_{(x,z) \in \mathcal{R} }   \frac{2L}{\frac{1}{nh^q}\sum_{k  \neq i} a_k^{(i)} } \ \right\} \cdot \max_{1 \leq i \leq n} \frac{ \|\hat{Z}_i - Z_i\|}{h^{q+1}}, 
\end{align*}
where $a_i^{(i)} \leq \zeta_i^{(x,z)} \leq a_i^{(i-1)}$ represents the mean value form of the remainder for each $(x,z) \in \mathcal{C}$.  We will now bound the first term in the last line of the previous display with high probability.  
Let $H(0) = B$.  Observe that:
\begin{align*}
& \min_{1 \leq i \leq n} \inf_{(x,z) \in \mathcal{R}} \frac{1}{nh^q} \sum_{k \neq i} a_k^{(i)}  \\ 
\geq \ & \   \min_{1 \leq i \leq n} \inf_{(x,z) \in \mathcal{R}} \frac{1}{nh^q} \sum_{k \neq i} K\left(\frac{\| X_k - x\| + \|Z_k - z \|}{h}\right) -  \max_{1 \leq i \leq n} \frac{ L \|\hat{Z}_i - Z\|  }{h^{q+1}} \\
\geq  \ & \  \inf_{(x,z) \in \mathcal{R}} \frac{1}{nh^q} \sum_{k=1}^n K\left(\frac{\| X_k - x\| + \|Z_k - z \|}{h}\right) - \frac{B}{nh^q} -  \max_{1 \leq i \leq n} \frac{ L \|\hat{Z}_i - Z\|  }{h^{q+1}}. 
\end{align*}
The latter two terms converge to zero by assumption.

Now, let $\hat{f}_{XZ}(x,z) = \frac{1}{nh^q} \sum_{k=1}^n K\left(\frac{\| X_k - x\| + \|Z_k - z \|}{h}\right)$ and $\bar{f}_{XZ}(x,z) = \mathbb{E}[\hat{f}_{XZ}(x,z)]$.  

 For the first term, we decompose further as:
\begin{align*}
 & \inf_{(x,z) \in \mathcal{R}} \frac{1}{nh^q} \sum_{k=1}^n K\left(\frac{\| X_k - x\| + \|Z_k - z \|}{h}\right) \\ 
 \geq \ & \  \inf_{(x,z) \in \mathcal{R}} \bar{f}_{XZ}(x,z) - \sup_{(x,z) \in \mathcal{R}} |\hat{f}_{XZ}(x,z)- \bar{f}_{XZ}(x,z)|.
\end{align*}
It is well known that kernel density estimation suffers from boundary bias and is inconsistent near the boundary when the density is non-zero; therefore, we cannot expect a bias term $\sup_{(x,z) \in \mathcal{R}} |\bar{f}_{XZ}(x,z) - f_{XZ}(x,z) |$ to converge to 0.  Nevertheless, we can lower bound the smoothed density $\bar{f}_{XZ}(x,z)$ directly. Observe that, for $h_n$ small enough,
\begin{align*}
\inf_{(x,z) \in \mathcal{R}} \bar{f}_{XZ}(x,z) & \geq \frac{k}{h_n^q} \inf_{(x,z) \in \mathcal{R}} P(\|(X-x,Z-z) \|_2 \leq rh)
\\ & \geq \frac{k}{h_n^q}  \inf_{(x,z) \in \mathcal{R}} P(\sqrt{q}\|(X-x,Z-z) \|_\infty \leq rh) 
\\ & \geq \frac{\delta k r^q}{q^{q/2}}.
\end{align*}
For the last line, observe that $L_\infty$-balls with center in $\mathcal{R}$ and radius $rh/\sqrt{q}$ that have the most volume outside of $\mathcal{R}$ are those centered at the vertices of $\mathcal{R}$.  At these points, it can readily seen that if $h$ is small enough half of each edge is contained in $\mathcal{R}$.  Therefore, for these extremal hyperrectangles, the region inside $\mathcal{R}$ is itself a hyperrectangle, with volume $(rh/ \sqrt{q})^q$. Thus, for any such $L_\infty$ ball with center in $\mathcal{R}$, the probability that $(X,Z)$ is contained in the ball is lower bounded by $\delta(rh/\sqrt{q})^q$.



For the final term, Theorem 2.3 of \citet{AIHPB_2002__38_6_907_0} implies that this is $o_P(1)$.

Now we verify the input stability condition.  Observe that:
\begin{align*}
& \max_{1 \leq i \leq n+1} \left| \hat{F}_n(Y_{n+i} \ | \ X_{n+i}, \hat{Z}_{n+i}) - \hat{F}_n(Y_{n+i} \ | \ X_{n+i}, Z_{n+i}) \right| \\ 
\leq & \max_{1 \leq i \leq n+1} \left| \underbrace{\frac{\sum_{k=1}^n K\left(\frac{\|X_k-X_{n+i}\| + \|\hat{Z}_k-Z_{n+i}\|}{h} \right) }{\sum_{k=1}^n K\left(\frac{\|X_k-X_{n+i}\| + \|\hat{Z}_k-\hat{Z}_{n+i}\|}{h} \right)}}_{A_i} \cdot \underbrace{\frac{\sum_{j=1}^n K\left(\frac{\|X_j-X_{n+i}\| + \|\hat{Z}_j-\hat{Z}_{n+i}\|}{h} \right) \mathbbm{1}(Y_j \leq y)}{\sum_{k=1}^n K\left(\frac{\|X_k-X_{n+i}\| + \|\hat{Z}_k-Z_{n+i}\|}{h} \right) }}_{B_i}   \right.
\\  & \ \ \ \ \ \ \ \ \ \  - \left.  \underbrace{\frac{\sum_{j=1}^n K\left(\frac{\|X_j-X_{n+i}\| + \| \hat{Z}_j-Z_{n+i}\|}{h} \right) \mathbbm{1}(Y_j \leq y)}{\sum_{k=1}^n K\left(\frac{\|X_k-X_{n+i}\| + \|\hat{Z}_k-Z_{n+i}\|}{h} \right) }}_{C_i} \  \right| 
\\ & \leq \max_{1 \leq i \leq n+1} \left| A_i-1 \right| \cdot \left| B_i \right| + \max_{1 \leq i \leq n+1}  \left|B_i- C_i \right|, \text{ say. }
\end{align*}
We will first bound $\max_{1 \leq i \leq n+1} |A_i-1|$.  Observe that
\begin{align*}
 \max_{1 \leq i \leq n+1} (A_i-1) &  \leq \frac{1}{1 - \max_{1 \leq i \leq n+1} R_i},  
\end{align*}
where $ \max_{1 \leq i \leq n+1} R_i$ takes the form:
\begin{align*}
\ & \ \max_{1 \leq i \leq n+1} R_i 
\\ &= L \max_{1 \leq i \leq n+1} \frac{\|\hat{Z}_{n+i} - Z_{n+i} \|}{h^{q+1}} \cdot  \left(\frac{1}{nh^q}\sum_{k=1}^n K\left(\frac{\|X_k-X_{n+i}\| + \|\hat{Z}_k-Z_{n+i}\|}{h} \right) \right)^{-1} \\  
  & \leq L \max_{1 \leq i \leq n+1} \frac{\|\hat{Z}_{n+i} - Z_{n+i} \|}{h^{q+1}} \cdot  \left( \inf_{(x,z) \in \mathcal{R}} \frac{1}{nh^q}\sum_{k=1}^n K\left(\frac{\|X_k-x\| + \|Z_k-z\|}{h} \right) - \max_{1 \leq i \leq n} \frac{\|\hat{Z}_{i} - Z_{i} \|}{h^{q+1}} \right)^{-1}.  
\end{align*}
For the inverse term, we can repeat arguments used  to verify the input stability condition to conclude that the term is $O_P(1)$.  Since $|B_i| \leq 1+o_P(1)$ by similar arguments, we have that first term us upper bounded by a term that is $o_P(1)$.   The lower bound is analogous.

Now, for the second term, we have, repeating a similar argument to one used above, 
\begin{align*}
\ & \ \max_{1 \leq i \leq n+1} |B_i - C_i| 
 \\ & \leq \left\{L \max_{1 \leq i \leq n+1}  \frac{\| \hat{Z}_i - Z_i \| }{h^{q+1}} \cdot \left(\min_{1 \leq i \leq n+1} \frac{1}{nh^q} \sum_{j=1}^n K\left(\frac{\|X_j-X_{n+i}\| + \|\hat{Z}_j-Z_{n+i}\|}{h} \right) \right)^{-1} \right\}
 \\ &= o_P(1).
\end{align*}
Thus, the training stability condition is satisfied.  
 
Finally, for consistency of distribution-free regression, we slightly modify proof of \citet{10.1214/aos/1176346815}; we provide some additional details for completeness.  

Mirroring the notation of the above reference, let $m(x_1,z_1;y) = \mathbb{E}[\mathbbm{1}(Y \leq y) \ | X=x_1, Z=z_1]$ and define the expectation:
\begin{align*}
U_h(x_1,z_1;y) = \frac{\int H\left(\frac{\|x_2 - x_1 \| + \|z_2 - z_1 \| }{h}  \right) m(x_1, z_1;y) \ \mu(dx_2 \ dz_2)}{\int H\left(\frac{\|x_2 - x_1 \| + \|z_2 - z_1 \| }{h}  \right) \ \mu(dx_2 \ dz_2)}.
\end{align*}

Similar to Lemma 1 of the above reference, we will show that:
\begin{align*}
U_h(X_1,Z_1;Y) \xrightarrow{a.s.} m(X_1,Z_1;Y)
\end{align*}
Since $|U_h(x_1,z_1;y)| \leq 1$, the bounded convergence theorem would then imply that:
\begin{align*}
  \mathbb{E}|U_h(X_1,Z_1;Y) - m(X_1,Z_1;Y)| \rightarrow 0  
\end{align*}
It would then remain to show
$\mathbb{E}| \hat{m}(X_1,Z_1;Y) - U_h(X_1,Z_1;Y)| \rightarrow 0$. Now, observe that:
\begin{align*}
& |U_h(x_1,z_1;y) - m(x_1,z_1;y)|
\\ \leq  & \ \frac{\int H\left( \frac{\|(x_1,z_1) - (x_2,z_2 ) \|}{h} \right) \left|m(x_1,z_1;y) - m(x_2,z_2;y)  \right|  \mu(dx_2 \ dz_2)}{\int H\left( \frac{\|(x_1,z_1) - (x_2,z_2 ) \|}{h} \right) \ \mu(dx_2 \ dz_2)} \\ & \times \frac{\int H\left( \frac{\|(x_1,z_1) - (x_2,z_2 ) \|}{h} \right) \mu(dx_2 \ dz_2) } {\int H\left( \frac{\|x_1- x_2\| +\| z_1 - z_2 \|}{h} \right) \mu(dx_2 \ dz_2)}.
\end{align*}
Since $(X,Z)$ is compactly supported on a hyperrectangle, the denominator of the second term in the product is lower bounded by $H(\frac{D_1 + D_2}{h})$, where $D_1$ and $D_2$ are maximal distances between points in the support of $X$ and $Z$, respectively.  Therefore, the denominator of this term converges to $B$ as $h_n \rightarrow 0$, as does the numerator.  For the first term, the result follows directly from the proof of Lemma 1 of the above reference.   

Now, $\mathbb{E}| \hat{m}(X_1,Z_1;Y) - U_h(X_1,Z_1;Y)| \rightarrow 0$ can be shown via a straightforward modification of the variance calculation in the proof of Theorem 1 of the above reference. The result follows. \qed



\end{document}